\documentclass[english,12pt]{article}

\usepackage[T1]{fontenc}
\usepackage{textcomp}
\usepackage[utf8]{inputenc}
\usepackage{colortbl}
\usepackage{array}
\usepackage{multirow}
\usepackage{varwidth}
\usepackage{amsmath}
\usepackage{amsthm}
\usepackage{amssymb}
\usepackage{graphicx}
\usepackage[authoryear]{natbib}

\makeatletter

\providecommand{\tabularnewline}{\\}
\newenvironment{cellvarwidth}[1][t]
    {\begin{varwidth}[#1]{\linewidth}}
    {\@finalstrut\@arstrutbox\end{varwidth}}

\usepackage[letterpaper,margin=1.0in]{geometry}
\usepackage{hyperref}
\usepackage{bookmark}
\usepackage{enumitem}
\usepackage{threeparttable}
\usepackage{longtable}
\usepackage{lipsum}
\usepackage{array}
\allowdisplaybreaks
\usepackage[T1]{fontenc}
\usepackage{setspace}
\usepackage{etoolbox}
\makeatletter
\patchcmd{\NAT@citex}
  {\@citea\NAT@hyper@{%
     \NAT@nmfmt{\NAT@nm}%
     \hyper@natlinkbreak{\NAT@aysep\NAT@spacechar}{\@citeb\@extra@b@citeb}%
     \NAT@date}}
  {\@citea\NAT@nmfmt{\NAT@nm}%
   \NAT@aysep\NAT@spacechar\NAT@hyper@{\NAT@date}}{}{}
\patchcmd{\NAT@citex}
  {\@citea\NAT@hyper@{%
     \NAT@nmfmt{\NAT@nm}%
     \hyper@natlinkbreak{\NAT@spacechar\NAT@@open\if*#1*\else#1\NAT@spacechar\fi}%
       {\@citeb\@extra@b@citeb}%
     \NAT@date}}
  {\@citea\NAT@nmfmt{\NAT@nm}%
   \NAT@spacechar\NAT@@open\if*#1*\else#1\NAT@spacechar\fi\NAT@hyper@{\NAT@date}}
  {}{}
\makeatother

\makeatother

\theoremstyle{plain}
\newtheorem{assumption}{\protect\assumptionname}
\newtheorem{thm}{\protect\theoremname}
\usepackage{babel}
\providecommand{\assumptionname}{Assumption}
\providecommand{\theoremname}{Theorem}

\begin{document}
\title{Estimating Treatment Effects in Panel Data Without Parallel Trends}
\author{Shoya Ishimaru\footnote{Hitotsubashi University, Department of Economics (email: shoya.ishimaru@r.hit-u.ac.jp). \newline I thank Stéphane Bonhomme, Yu–Chang Chen, Mark Colas, Xavier D'Haultfoeuille, Joachim Freyberger, Hiroyuki Kasahara, Eric Klemm, Kenichi Nagasawa, and many seminar participants for helpful comments and suggestions. This study uses the factually anonymous Sample
of Integrated Labour Market Biographies (version 1975--2021). Data access was provided via a Scientific Use File supplied by the Research Data Centre (FDZ) of the German Federal Employment Agency (BA) at the Institute for Employment Research (IAB). Support from the Japan Society for Promotion of Science (Grant No. 21K13310) is gratefully acknowledged. All errors are mine.}}
\maketitle
\begin{abstract}
This paper proposes a novel approach for estimating treatment effects
in panel data settings, addressing key limitations of the standard
difference-in-differences (DID) approach. The standard approach relies
on the parallel trends assumption, implicitly requiring that unobservable
factors correlated with treatment assignment be unidimensional, time-invariant,
and affect untreated potential outcomes in an additively separable
manner. This paper introduces a more flexible framework that allows
for multidimensional unobservables and non-additive separability,
and provides sufficient conditions for identifying the average treatment
effect on the treated. An empirical application to job displacement
reveals substantially smaller long-run earnings losses compared to
the standard DID approach, demonstrating the framework's ability to
account for unobserved heterogeneity that manifests as differential
outcome trajectories between treated and control groups.\vfill
\end{abstract}

\section{Introduction}

The difference-in-differences (DID) approach is one of the most widely
used methods for estimating treatment effects in panel data, particularly
when treatment assignment is nonrandom. Its core identifying assumption,
parallel trends, is typically justified by an additively separable
model of untreated potential outcomes:
\begin{equation}
Y_{it}(0)=U_{i}+\mu_{t}+\varepsilon_{it},\label{eq:DID_model}
\end{equation}
where $U_{i}$ represents a unit-specific unobservable that may correlate
with treatment, $\mu_{t}$ captures time effects, and $\varepsilon_{it}$
is an idiosyncratic error uncorrelated with treatment. This additive
form is sufficient but not necessary for parallel trends. However,
nonseparable alternatives that support parallel trends are difficult
to construct without assuming either that $U_{i}$ is independent
of treatment or that untreated outcomes are stationary over time,\footnote{In either case, a single difference (cross-sectional or before--after)
already identifies the effect, making DID superfluous.} making this additive specification the implicit foundation of the
DID framework.

This paper relaxes this implicit restriction by developing a framework
that accommodates multidimensional unobservables and nonseparable
structures. I derive sufficient conditions for identifying treatment
effects, drawing on nonclassical measurement error models \citep{hu2008instrumental}.
The key insight is that repeated observations of untreated outcomes,
prevalent in many DID applications beyond the canonical two-period
design, act as multiple noisy measurements of underlying unobserved
heterogeneity. This approach enables identification of treatment effects
under nonparametric conditions without requiring parallel trends to
hold, providing a flexible alternative to the restrictive additive
framework. These conditions encompass a broad class of specifications,
including models with time-varying unobserved heterogeneity such as
hidden Markov models. A central requirement is that the dimension
of unobservables does not exceed the number of pre- or post-treatment
periods, making the approach most applicable in settings with rich
longitudinal data.

As an empirical illustration of my framework, I examine the impact
of job displacement on earnings. In the job displacement literature,
a central concern is unobserved heterogeneity and its influence on
earnings trajectories. The standard DID approach, by relying on the
parallel trends assumption, allows for differences in earnings levels
but requires identical trends across treatment and control groups.
This leads to biased estimates when the groups differ in their earnings
growth trajectories. By estimating a more flexible model that does
not require the parallel trends assumption, this paper provides an
alternative for estimating the impact of job displacement. The empirical
results indicate that this approach yields smaller estimated impacts
of job displacement on earnings than the standard DID method. This
difference is particularly pronounced in the long run, where the estimated
reduction in earnings nine years after displacement is approximately
half of that suggested by the standard DID estimate.

While convenient, the parallel trends assumption underlying the DID
framework often lacks a clear theoretical foundation and relies heavily
on empirical validation. Researchers typically check for pre-treatment
trends to assess its validity. However, if pre-treatment trends are
not parallel, options for correcting the model are limited. Even if
parallel trends appear to hold in the pre-treatment periods, this
does not guarantee that they would continue to hold in the post-treatment
periods. Furthermore, the additively separable model is vulnerable
to nonlinear transformations, making it difficult to justify parallel
trends across different transformations of the outcome variable \citep{roth2023parallel}.

Addressing potential deviations from parallel trends is therefore
a critical issue in the DID framework. Traditionally, empirical studies
often incorporate unit-specific trends in robustness checks (e.g.,
\citealp{jacobson1993earnings}), which, while introducing two-dimensional
unobservables, still maintain an additive structure similar to the
standard specification in equation (\ref{eq:DID_model}) and assume
linear trends without a clear theoretical foundation. \citet{athey2006identification}
relax additivity by allowing nonseparable models under a changes-in-changes
framework, though their approach requires that a unidimensional unobservable
be the sole determinant of untreated outcomes. A recent strand of
the literature employs linear factor models (i.e., $Y_{it}(0)=U_{i}'\gamma_{t}+\mu_{t}+\varepsilon_{it}$
with multidimensional $U_{i}$) for identification and estimation
of treatment effects (e.g., \citealp{imbens2021controlling,callawaykarami2023treatment,callaway2023treatment,brown2025dynamic}).
Another recent approach involves partial identification techniques
that extrapolate pre-treatment trends (\citealp{rambachan2023more}).
My approach provides point identification without imposing specific
functional forms, and achieves this through different identifying
restrictions.

DID analyses sometimes incorporate pre-treatment outcomes as covariates
(e.g., \citealp{heckman1998matching,smith2005does}). However, theoretical
foundations for this practice remain debated (\citealp{daw2018matching,park2024matching,chabe2025should}).
While pre-treatment outcomes may capture unobserved heterogeneity
that differencing alone does not eliminate, the transitory shocks
they inherently contain act as measurement error, limiting their suitability
as covariates. My framework provides a theoretically grounded alternative
to this practice by explicitly considering pre-treatment outcomes
as noisy measurements of underlying unobserved heterogeneity. The
framework establishes a coherent foundation for how pre-treatment
outcomes can be leveraged to strengthen causal inference in panel
data settings.

This paper also relates to the literature on latent variable methods
in econometrics. \citet{hu2008instrumental} develop foundational
techniques for identification in nonclassical measurement error models,
showing how completeness conditions enable identification of economic
and econometric models with unobservables. These completeness-based
methods have been developed and applied in various contexts, including
estimation of skill formation models (\citealp{cunha2010estimating}),
identification in nonparametric fixed effects models (\citealp{deaner2023proxy}),
and treatment effect estimation with mismeasured covariates (\citealp{nagasawa2022treatmenteffectestimationnoisy,deaner2023controlling}).
While addressing different empirical problems, these studies share
with the present paper the core insight that repeated measurements
can be used to identify causal relationships in the presence of unobserved
heterogeneity.

\section{Econometric Framework\label{sec:Econometric-Framework}}

This section provides a general framework for identifying treatment
effects without relying on the parallel trends assumption. Section
\ref{subsec:Setup-and-Assumptions} describes essential assumptions
governing the relationship between potential outcomes, treatment status,
and unobservables. Section \ref{subsec:Examples} illustrates specific
econometric models that adhere to these assumptions, showcasing the
applicability and flexibility of the framework. Section \ref{subsec:Identification}
establishes the identification of treatment effect parameters under
the outlined conditions. Section \ref{subsec:Estimation} proposes
the estimation strategy.

\subsection{Setup and Assumptions\label{subsec:Setup-and-Assumptions}}

I consider a nonstaggered DID setting with a balanced panel. Units
are indexed by $i\in\{1,\ldots,N\}$ and periods by $t\in\{-T_{\text{pre}},\ldots,T_{\text{post}}\}$.
Each unit is classified into either a treatment group ($D_{i}=1$)
or a control group ($D_{i}=0$). Units in the treatment group receive
the treatment beginning in period $t=1$. I refer to periods $t\in\{-T_{\text{pre}},\ldots,-1\}$
as pre-treatment periods, $t=0$ as the reference period, and $t\in\{1,\ldots,T_{\text{post}}\}$
as post-treatment periods. I assume random variables are i.i.d. across
units but not necessarily across periods, adhering to a large-$N$,
fixed-$T$ framework. For notational simplicity, I omit the unit index
$i$ from density functions, as densities are identical across units
under the i.i.d. assumption.

Let $Y_{it}(d)$ represent a potential outcome in period $t$ with
treatment status $d\in\{0,1\}$. As the treatment group receives the
treatment in post-treatment periods, the observed outcome is given
by:
\begin{equation}
Y_{it}=\begin{cases}
Y_{it}(0) & \text{if }t\in\{-T_{\text{pre}},\ldots,0\}\\
D_{i}Y_{it}(1)+(1-D_{i})Y_{it}(0) & \text{if }t\in\{1,\ldots,T_{\text{post}}\}
\end{cases}
\end{equation}
The potential outcomes $\left\{ Y_{it}(0),Y_{it}(1)\right\} ^{T_{\text{post}}}_{t=-T_{\text{pre}}}$
and the treatment group membership $D_{i}$ may depend on a $K$-dimensional
unobservable $U_{i}$ distributed on the support ${\cal U}\subset\mathbb{R}^{K}$.
Observed covariates are allowed but are kept implicit; all assumptions
are conditioned on these covariates. I denote untreated outcomes in
pre-treatment periods by 
\begin{equation}
\boldsymbol{Y}^{\text{pre}}_{i}(0)\equiv\left(Y_{i,-T_{\text{pre}}}(0),\ldots,Y_{i,-1}(0)\right),\label{eq:Y_pre}
\end{equation}
 and untreated outcomes in post-treatment periods by 
\begin{equation}
\boldsymbol{Y}^{\text{post}}_{i}(0)\equiv\left(Y_{i,1}(0),\ldots,Y_{i,T_{\text{post}}}(0)\right).\label{eq:Y_post}
\end{equation}

I focus on identifying the average treatment effect on the treated
(ATT), 
\begin{equation}
\theta^{\text{ATT}}_{t}\equiv E\left[Y_{it}(1)-Y_{it}(0)|D_{i}=1\right]\text{ for }t\in\{1,\ldots,T_{\text{post}}\},
\end{equation}
which is a target parameter in a standard DID setting. The standard
setting assumes parallel trends for untreated outcomes
\begin{equation}
E\left[Y_{it}(0)-Y_{i0}(0)|D_{i}=1\right]=E\left[Y_{it}(0)-Y_{i0}(0)|D_{i}=0\right]\text{ for }t\in\{1,\ldots,T_{\text{post}}\},\label{eq:PT_assumption}
\end{equation}
which makes the following statistical parameter identical to $\theta^{\text{ATT}}_{t}$:
\begin{equation}
\theta^{\text{DID}}_{t}\equiv E\left[Y_{it}-Y_{i0}|D_{i}=1\right]-E\left[Y_{it}-Y_{i0}|D_{i}=0\right].\label{eq:DID_parameter}
\end{equation}

In the following, I establish general conditions that enable the identification
of the ATT without the parallel trends assumption in equation (\ref{eq:PT_assumption}).
Instead of explicitly specifying the model of untreated outcomes,
I express assumptions in terms of the joint distribution of untreated
outcomes $\left\{ Y_{it}(0)\right\} ^{T_{\text{post}}}_{t=-T_{\text{pre}}}$,
treatment group membership $D_{i}$, and the unobservable $U_{i}$.
No functional form or parametric assumptions are made on the relationship
between $Y_{it}(0)$ and $U_{i}$, unlike in the additively separable
model (\ref{eq:DID_model}) underlying the standard DID framework.
As for treated outcomes $\left\{ Y_{it}(1)\right\} ^{T_{\text{post}}}_{t=1}$,
I follow the standard approach and do not impose any assumptions on
them, allowing for arbitrary treatment effect heterogeneity. Examples
of specific models that satisfy these conditions are provided in Section
\ref{subsec:Examples}.
\begin{assumption}[Conditional Independence]

$\boldsymbol{Y}^{\text{pre}}_{i}(0)$, $\left(Y_{i0}(0),D_{i}\right)$,
and $\boldsymbol{Y}^{\text{post}}_{i}(0)$ are mutually independent
conditional on $U_{i}$.
\end{assumption}
Assumption 1 can be compared to the standard DID framework based on
equation (\ref{eq:DID_model}) along three dimensions. First, it requires
untreated outcomes to be correlated with treatment only through $U_{i}$,
as in standard DID. Second, it relaxes a key DID restriction by allowing
arbitrary correlation between the reference period outcome $Y_{i0}(0)$
and treatment $D_{i}$, thereby accommodating ``Ashenfelter's dip.''\footnote{In a DID application, \citet{ashenfelter1978estimating} documents
what is now called ``Ashenfelter's dip,'' a decline in earnings
prior to participation in job training. Within the standard framework,
this issue is typically addressed by shifting the reference period
to precede the dip.} Third, it imposes an additional restriction on serial correlation
of untreated outcomes: the dependence across $\boldsymbol{Y}^{\text{pre}}_{i}(0)$,
$Y_{i0}(0)$, and $\boldsymbol{Y}^{\text{post}}_{i}(0)$ must be solely
through $U_{i}$, although arbitrary dependence within $\boldsymbol{Y}^{\text{pre}}_{i}(0)$
or within $\boldsymbol{Y}^{\text{post}}_{i}(0)$ remains allowed.
This third point is the only one that comes at a cost relative to
standard DID. Its plausibility can be strengthened by separating the
timing of pre-treatment observations ($t\le-1$) and post-treatment
observations ($t\ge1$) from the reference period ($t=0$), or by
absorbing persistent shocks into $U_{i}$, as illustrated in Section
\ref{subsec:Examples} by a hidden Markov model.
\begin{assumption}[Nondeterministic Treatment]

The treatment assignment $D_{i}$ satisfies $0<\Pr[D_{i}=1|U_{i}]<1$
almost surely.
\end{assumption}
Assumption 2 requires that treatment assignment retain an element
of randomness conditional on $U_{i}$, ensuring that both treatment
and control groups can be observed for any given value of $U_{i}$.
It follows from this assumption that $U_{i}|D_{i}=1$ and $U_{i}|D_{i}=0$
have the same support, a necessary condition for identifying the ATT
through comparisons across treatment statuses.
\begin{assumption}[Hu and Schennach (2008) Conditions--I]
\,

\begin{enumerate}[label=(\alph*)]

\item The joint, marginal, and conditional densities of $(\boldsymbol{Y}^{\text{pre}}_{i}(0),Y_{i0}(0),\boldsymbol{Y}^{\text{post}}_{i}(0),U_{i})$
given $D_{i}=d$ are all bounded for $d\in\{0,1\}$. 

\item If $\int_{u\in{\cal U}}f_{\boldsymbol{Y}^{\text{pre}}(0)|U}(y|u)g(u)du=0$
for any $y\in\mathbb{R}^{T_{\text{pre}}}$ and $\int_{u\in{\cal U}}|g(u)|du<\infty$,
then $g=0$.

\end{enumerate}
\end{assumption}
Assumption 3 follows \citet{hu2008instrumental}. Condition (a) requires
bounded densities, and condition (b) requires completeness of $f_{\boldsymbol{Y}^{\text{pre}}(0)|U}$.
Sufficient conditions for completeness in a variety of special cases
have been provided in the literature (e.g., \citealp{newey2003instrumental,d2011completeness,andrews2017examples,hu2018nonparametric}),
but no simple general rule characterizes when completeness holds.
It is therefore useful to build intuition through the lens of a measurement
error problem. The observed distribution satisfies:
\begin{equation}
f_{\boldsymbol{Y}^{\text{pre}}(0)}(y)=\int_{u\in{\cal U}}f_{\boldsymbol{Y}^{\text{pre}}(0)|U}(y|u)f_{U}(u)du.
\end{equation}
Here, $\boldsymbol{Y}^{\text{pre}}_{i}(0)$ can be seen as a noisy
measurement of $U_{i}$, with $f_{\boldsymbol{Y}^{\text{pre}}(0)|U}$
being the distribution of measurement error. Completeness of $f_{\boldsymbol{Y}^{\text{pre}}(0)|U}$
implies that this integral equation has a unique solution for $f_{U}$
when $f_{\boldsymbol{Y}^{\text{pre}}(0)}$ and $f_{\boldsymbol{Y}^{\text{pre}}(0)|U}$
are provided. Put differently, the distribution of unobserved heterogeneity
must be recoverable from the observed outcomes and the way they are
distorted by noise. This requirement rules out cases where $\boldsymbol{Y}^{\text{pre}}_{i}(0)$
acts as a ``compressed'' measurement of $U_{i}$. For example, if
the dimension of $U_{i}$ exceeds that of $\boldsymbol{Y}^{\text{pre}}_{i}(0)$
(i.e., $K>T_{\text{pre}}$) or $U_{i}$ affects $\boldsymbol{Y}^{\text{pre}}_{i}(0)$
only through a lower-dimensional index $g(U_{i})$, then recovery
would fail because the measurement does not contain enough independent
information about the underlying unobserved heterogeneity. Completeness
is the nonparametric counterpart of the rank condition in linear models,
a parallel drawn by \citet{newey2003instrumental} in the instrumental
variable setting. As with the rank condition, the comparison of dimensions
indicates when completeness is plausible in general rather than characterizing
exactly when it holds.
\begin{assumption}[Hu and Schennach (2008) Conditions--II]
\,

\begin{enumerate}[label=(\alph*)]

\item If $\int_{u\in{\cal U}}f_{\boldsymbol{Y}^{\text{post}}(0)|U}(y|u)g(u)du=0$
for any $y\in\mathbb{R}^{T_{\text{post}}}$ and $\int_{u\in{\cal U}}|g(u)|du<\infty$,
then $g=0$.

\item $U_{i}$ admits a normalization using a known functional $M$
such that $M\left[f_{\boldsymbol{Y}^{\text{pre}}(0)|U}(\cdot|u)\right]=u$
for any $u\in{\cal U}$.

\item $\int\mathbb{I}\left\{ f_{Y_{0}(0)|U,D=0}\left(y|u_{1}\right)\ne f_{Y_{0}(0)|U,D=0}\left(y|u_{2}\right)\right\} f_{Y_{0}(0)|D=0}(y)dy>0$
for any $u_{1},u_{2}\in{\cal U}$ with $u_{1}\ne u_{2}$.

\end{enumerate}
\end{assumption}
Assumption 4 also follows \citet{hu2008instrumental} and provides
one pathway to identification. Condition (a) requires completeness
of $f_{\boldsymbol{Y}^{\text{post}}(0)|U}$, analogous to Assumption
3--(b). Together, these two completeness conditions lead to a requirement
$K\le\min\{T_{\text{pre}},T_{\text{post}}\}$ in general. Condition
(b) normalizes $U_{i}$ based on $\boldsymbol{Y}^{\text{pre}}_{i}(0)$.\footnote{\citet{deaner2023controlling} shows that the \citet{hu2008instrumental}
identification result holds up to a bijective transformation of the
latent variable. Since the treatment effect parameters are invariant
to such transformations, this normalization assumption is not strictly
necessary. I retain it here to simplify the identification argument.} Finally, condition (c) specifies a requirement for $Y_{i0}(0)$,
which complements $\boldsymbol{Y}^{\text{pre}}_{i}(0)$ and $\boldsymbol{Y}^{\text{post}}_{i}(0)$
as a third measurement of $U_{i}$.\footnote{\citet{Hu2017unobs} describes this identification strategy as a ``2.1-measurement
model,'' where ``2.1'' indicates that the third measurement can
be as simple as a binary indicator.} This requirement is weaker than the completeness conditions imposed
on the other measurements. The condition requires that for any two
distinct values of $U_{i}$, the corresponding conditional densities
of $Y_{i0}(0)$ differ on a set of positive measure. In addition to
requiring $Y_{i0}(0)$ to depend on each dimension of $U_{i}$ in
some way, this requirement rules out cases where the dependence on
$U_{i}$ could be reduced to a lower-dimensional index. For example,
if a lower-dimensional index $g_{0}(U_{i})$ serves as a sufficient
statistic for the conditional distribution of $Y_{i0}(0)$, then $f_{Y_{0}(0)|U,D=0}\left(y|u_{1}\right)=f_{Y_{0}(0)|U,D=0}\left(y|u_{2}\right)$
would hold for $u_{1}\ne u_{2}$ with $g_{0}(u_{1})=g_{0}(u_{2})$,
violating condition (c).

When $K\ge2$, condition (c) is incompatible with single-index specifications,
where the unobservable affects the outcome through a single distributional
feature such as a location shifter. While such specifications impose
a dimension reduction that the data may not truly satisfy, they are
commonly used in practice for tractability. Yet the very functional
form restrictions these specifications impose often provide alternative
leverage for identification. The following assumption applies instead
of Assumption 4 when such alternative pathways are available.
\begin{assumption}[Direct Identification]

The distributions of $\boldsymbol{Y}^{\text{pre}}_{i}(0)|U_{i},D_{i}=0$
and $\boldsymbol{Y}^{\text{post}}_{i}(0)|\allowbreak U_{i},D_{i}=0$
are identified from the joint distribution of $\left(\boldsymbol{Y}^{\text{pre}}_{i}(0),Y_{i0}(0),\boldsymbol{Y}^{\text{post}}_{i}(0)\right)$
conditional on $D_{i}=0$.
\end{assumption}
Assumption 5 directly assumes identification of conditional distributions
from the control group data. Assumptions 1, 3, and 4 together imply
this assumption, building on \citet{hu2008instrumental}. It can be
established through several alternative pathways, including the identification
strategy of \citet{freyberger2018non} for linear and nonlinear factor
models and strategies in parametric earnings models (e.g., \citealp{guvenen2007learning}).
While inconsistent with Assumption 4--(c) due to their single-index
structure, these models achieve identification by exploiting their
functional form restrictions, typically with the same data requirement
$K\le\min\{T_{\text{pre}},T_{\text{post}}\}$ as the pathway through
Assumption 4.

\subsection{Examples\label{subsec:Examples}}

I now provide specific examples illustrating how various well-known
econometric models fit within the general framework established in
Section \ref{subsec:Setup-and-Assumptions}.

\subsubsection{Standard DID}

The first example involves the standard DID setup, which can be adapted
to the framework in Section \ref{subsec:Setup-and-Assumptions} by
modifying its distributional assumptions. The parallel trends condition
implicitly rests on an additively separable model of untreated potential
outcomes in equation (\ref{eq:DID_model}). In the standard DID approach,
mean independence is assumed, $E\left[\varepsilon_{it}|D_{i}\right]=0$
for all $t$. In my framework, I relax this assumption for $t=0$
but impose (distributional) independence of $\varepsilon_{it}$ from
$D_{i}$ for $t\ne0$ given $(U_{i},\varepsilon_{i0})$. The standard
DID approach allows arbitrary serial dependence of error terms $\left\{ \varepsilon_{it}\right\} ^{T_{\text{post}}}_{t=-T_{\text{pre}}}$.
Within my framework, Assumption 1 requires that $\varepsilon_{i0}$,
$\left\{ \varepsilon_{it}\right\} ^{-1}_{t=-T_{\text{pre}}}$, and
$\left\{ \varepsilon_{it}\right\} ^{T_{\text{post}}}_{t=1}$ be independent
from each other given $U_{i}$, although the dependence within $\left\{ \varepsilon_{it}\right\} ^{-1}_{t=-T_{\text{pre}}}$
or within $\left\{ \varepsilon_{it}\right\} ^{T_{\text{post}}}_{t=1}$
is still allowed. Beyond these differences in independence assumptions,
another distinction appears in robustness to nonlinear transformations:
while parallel trends typically fail after such transformations, the
identification conditions in my framework remain valid.

\subsubsection{Linear and Nonlinear Factor Models}

The second example involves factor models, which satisfy the assumptions
in Section \ref{subsec:Setup-and-Assumptions} when appropriate conditions
are imposed on the treatment assignment process. A linear factor model
is typically specified as:
\begin{equation}
Y_{it}(0)=U_{i}'\gamma_{t}+\mu_{t}+\varepsilon_{it},
\end{equation}
where $U_{i}$ is a $K$-dimensional vector of unobservables, $\gamma_{t}$
is a vector of factor loadings, and $\varepsilon_{it}$ is independent
across periods and independent of $U_{i}$. A special case of this
model is the unit-specific linear trend specification, where $U_{i}=(\alpha_{i},\beta_{i})'$
and $\gamma_{t}=(1,t)'$. Nonlinear extensions of the linear factor
model, such as those studied by \citet{freyberger2018non}, take the
form:
\begin{equation}
Y_{it}(0)=g_{t}\left(U_{i}'\gamma_{t}+\varepsilon_{it}\right),
\end{equation}
where $g_{t}$ is a period-specific nonlinear transformation. 

Under Assumption 1, the treatment status $D_{i}$ can depend on $(U_{i},\varepsilon_{i0})$
but must be independent of $\left\{ \varepsilon_{it}\right\} _{t\ne0}$
given $(U_{i},\varepsilon_{i0})$. Assumption 3--(b) holds as long
as factor loadings $\{\gamma_{t}\}^{-1}_{t=-T_{\text{pre}}}$ satisfy
a full rank condition and each $\varepsilon_{it}$ has a nonvanishing
characteristic function. However, these specifications do not satisfy
Assumption 4--(c), given that $U_{i}'\gamma_{0}$ serves as a single-index
sufficient statistic for the conditional distribution of $Y_{i0}(0)$.
This issue can be addressed by exploiting the serial independence
of $\varepsilon_{it}$ as in \citet{freyberger2018non} to invoke
Assumption 5.

The reliance on serial independence of $\varepsilon_{it}$ in this
example contrasts with recent treatment effect identification approaches
under linear factor models, which typically allow arbitrary serial
correlation of $\varepsilon_{it}$ (e.g., \citealp{callawaykarami2023treatment,callaway2023treatment,brown2025dynamic}).
The two frameworks are therefore not nested: my framework accommodates
nonseparable and nonlinear structures ruled out by linear factor models,
while those models accommodate serial correlation patterns ruled out
by Assumption 1. Within my framework, short-range dependence of moving-average
type can be accommodated by separating the timing of pre-, reference,
and post-treatment observations, at the cost of a longer panel. Persistent
dependence can be absorbed into $U_{i}$, as in the hidden Markov
example below, at the cost of a higher-dimensional unobservable.

\subsubsection{Hidden Markov Model}

The third example is a hidden Markov model, which also fits within
the framework in Section \ref{subsec:Setup-and-Assumptions} given
suitable assumptions on the treatment assignment process. \citet{arellano2017earnings}
employ the following model to study earnings dynamics:
\begin{equation}
Y_{it}(0)=\eta_{it}+\varepsilon_{it},\label{eq:hidden_Markov}
\end{equation}
where $\eta_{it}$ is a persistent earnings shock that follows a first-order
Markov process, modeled nonparametrically as $\eta_{it}=Q_{t}(\eta_{i,t-1},v_{it})$,
and $\varepsilon_{it}$ is a transitory shock. This model can be interpreted
within the framework by setting $U_{i}=\eta_{i0}$, which makes $\boldsymbol{Y}^{\text{pre}}_{i}(0)$,
$Y_{i0}(0)$, and $\boldsymbol{Y}^{\text{post}}_{i}(0)$ mutually
independent conditional on $U_{i}$ under a suitable independence
assumption on the transitory shocks. The treatment status $D_{i}$
can depend on $(\eta_{i0},\varepsilon_{i0})$, while $D_{i}$ must
be independent of $\left\{ \varepsilon_{it},\eta_{it}\right\} _{t\ne0}$
given $(\eta_{i0},\varepsilon_{i0})$.

\citet{arellano2017earnings} also consider a specification with permanent
unobserved heterogeneity:
\begin{equation}
Y_{it}(0)=\eta_{it}+\zeta_{i}+\varepsilon_{it},\label{eq:ABB-perm}
\end{equation}
where $\zeta_{i}$ represents a time-invariant individual effect.
This specification fits into the framework by setting $U_{i}=(\eta_{i0},\zeta_{i})$.
However, Assumption 4--(c) requires additional structure: the conditional
distribution of the transitory shock $\varepsilon_{i0}$ must depend
on $(\eta_{i0},\zeta_{i})$ in a manner that is not reducible to the
single index $\eta_{i0}+\zeta_{i}$, for instance through conditional
heteroskedasticity. Without such dependence, $\eta_{i0}+\zeta_{i}$
would serve as a sufficient statistic for the conditional distribution
of $Y_{i0}(0)$, violating Assumption 4--(c).

\subsubsection{Nonparametric Fixed Effects}

A fourth example is the nonparametric fixed effects model of \citet{deaner2023proxy},
which identifies causal effects by treating observations from different
periods as noisy proxies for unobserved heterogeneity. A static version
of this model specifies potential outcomes as
\begin{equation}
Y_{it}(d)=g_{dt}(U_{i},\varepsilon_{it}),\label{eq:nonpara-FE}
\end{equation}
where $g_{dt}$ is a nonseparable, period- and treatment-specific
function and $\varepsilon_{it}$ is independent across periods conditional
on $U_{i}$. This model also fits within the framework in Section
\ref{subsec:Setup-and-Assumptions} under suitable assumptions on
treatment assignment.

The identification strategy of \citet{deaner2023proxy} differs from
the one adopted here and cannot be directly applied to a DID design.
It builds on the approach of \citet{miao2018identifying}, whereas
the present framework follows \citet{hu2008instrumental}; \citet{deaner2023controlling}
discusses the distinction between \citet{hu2008instrumental} and
\citet{miao2018identifying}, and shows that the two frameworks are
not nested. Moreover, \citet{deaner2023proxy} identifies the period-$t$
effect by conditioning on treatment status in the immediately preceding
period $t-1$, leaving variation only in the first post-treatment
period in a DID design. Appendix \ref{sec:Adapting-the-Double-Proxy}
shows that the approach of \citet{miao2018identifying} can nonetheless
be adapted to a DID design, but only when the unobserved heterogeneity
is time-invariant.

\subsection{Identification\label{subsec:Identification}}

I now present the main identification result. The following theorem
establishes that the ATT parameters $\{\theta^{\text{ATT}}_{t}\}^{T_{\text{post}}}_{t=1}$
are identifiable from the observed data under the assumptions laid
out in Section \ref{subsec:Setup-and-Assumptions}.
\begin{thm}
Under Assumptions 1--3 and either Assumption 4 or 5, the ATT parameters
$\{\theta^{\text{ATT}}_{t}\}^{T_{\text{post}}}_{t=1}$ are identified
from the joint distribution of $\left(D_{i},Y_{i,-T_{\text{pre}}},\ldots,Y_{i,T_{\text{post}}}\right)$.
\end{thm}
The proof proceeds in three steps. The first step involves identifying
the distributions of $\boldsymbol{Y}^{\text{pre}}_{i}(0)|U_{i}$ and
$\boldsymbol{Y}^{\text{post}}_{i}(0)|U_{i}$ using control group data.
Given the definition of the observed outcome, $Y_{it}(0)$ is observed
for the control group ($D_{i}=0$) for all $t$. Assumption 1 implies
that $\boldsymbol{Y}^{\text{pre}}_{i}(0)$, $Y_{i0}(0)$, and $\boldsymbol{Y}^{\text{post}}_{i}(0)$
are mutually independent given $U_{i}$ and $D_{i}=0$. Combining
this with Assumptions 3 and 4, the distributions of $\boldsymbol{Y}^{\text{pre}}_{i}(0)|U_{i},D_{i}=0$
and $\boldsymbol{Y}^{\text{post}}_{i}(0)|U_{i},D_{i}=0$ can be identified
using the results from \citet{hu2008instrumental}. Alternatively,
Assumption 5 can be invoked when these distributions are identified
through other pathways (e.g., \citealp{freyberger2018non}). Assumption
1 implies that these distributions are identical to the distributions
of $\boldsymbol{Y}^{\text{pre}}_{i}(0)|U_{i}$ and $\boldsymbol{Y}^{\text{post}}_{i}(0)|U_{i}$,
and Assumption 2 ensures that they are identified over the entire
support of $U_{i}$.

The second step identifies the distribution of $U_{i}|D_{i}$. Given
Assumption 1, we have:
\begin{equation}
f_{\boldsymbol{Y}^{\text{pre}}(0)|D}(y|d)=\int_{u\in{\cal U}}f_{\boldsymbol{Y}^{\text{pre}}(0)|U}(y|u)f_{U|D}(u|d)du,
\end{equation}
for each $d\in\{0,1\}$. On the left-hand side, $f_{\boldsymbol{Y}^{\text{pre}}(0)|D}$
is directly available from the data. On the right-hand side, $f_{\boldsymbol{Y}^{\text{pre}}(0)|U}$
has been identified in the first step. Assumption 3 ensures that $f_{U|D}$
is uniquely determined through inversion.

Finally, the third step identifies the ATT parameters. Under Assumptions
1 and 2, the ATT parameter for each $t\in\{1,\ldots,T_{\text{post}}\}$
is characterized by the equation: 
\begin{equation}
\theta^{\text{ATT}}_{t}=\left\{ E\left[Y_{it}|D_{i}=1\right]-E\left[Y_{it}|D_{i}=0\right]\right\} -\int_{u\in{\cal U}}E\left[Y_{it}(0)|U_{i}=u\right]\left(f_{U|D=1}(u)-f_{U|D=0}(u)\right)du.\label{eq:ATT_equation}
\end{equation}
The first term represents the mean outcome difference between treatment
and control groups, which is directly available from the data. The
second term captures selection on unobservables, which depends on
distributions recovered in earlier steps.

Equation (\ref{eq:ATT_equation}) highlights a key distinction from
the standard DID framework in equation (\ref{eq:DID_parameter}).
The DID framework assumes that the selection bias term is constant
over time through its implicit additively separable structure, and
substitutes this term with $E\left[Y_{i0}|D_{i}=1\right]-E\left[Y_{i0}|D_{i}=0\right]$.
My approach offers a more flexible correction for selection bias by
identifying this term without functional form assumptions. 

\subsection{Estimation\label{subsec:Estimation}}

Given the identification result, estimation of the ATT parameters
$\{\theta^{\text{ATT}}_{t}\}^{T_{\text{post}}}_{t=1}$ can be implemented
in the following two steps. The first step estimates a parametric,
semiparametric, or nonparametric model using maximum likelihood based
on the joint distribution of observed untreated outcomes and treatment
assignment. The log-likelihood function is:
\begin{equation}
\sum^{N}_{i=1}\ln\int_{u\in{\cal U}}f_{\boldsymbol{Y}^{\text{pre}}(0)|U}\left(\boldsymbol{Y}^{\text{pre}}_{i}|u\right)f_{Y_{0}(0),D|U}\left(Y_{i0},D_{i}|u\right)f_{\boldsymbol{Y}^{\text{post}}(0)|U}\left(\boldsymbol{Y}^{\text{post}}_{i}|u\right)^{1-D_{i}}f_{U}(u)du,
\end{equation}
where the density functions can be parameterized by a finite-dimensional
parameter vector or represented using sieve approximations. Standard
asymptotic theory applies for the parametric case, while the sieve
case follows \citet{hu2008instrumental}. In the second step, plugging
the estimated distributions of $Y_{it}(0)|U_{i}$ and $U_{i}|D_{i}$
into equation (\ref{eq:ATT_equation}) yields the estimate of $\theta^{\text{ATT}}_{t}$
for each $t\in\{1,\ldots,T_{\text{post}}\}$. For standard errors,
\citet{ackerberg2012practical} show that a two-step estimator using
a series approximation in the first step can be treated as a parametric
estimator; Appendix \ref{subsec:ATT-Estimator-and} gives the asymptotic
expansion used in the empirical application.

While the identification result accommodates general heterogeneity
with dimension $K$ no greater than $\min\{T_{\text{pre}},T_{\text{post}}\}$,
estimation becomes increasingly complex as the dimension $K$ grows.
A fully flexible approach would require modeling conditional densities
of $T_{\text{pre}}$- and $T_{\text{post}}$-dimensional outcomes
given $K$-dimensional unobservables. Nonparametric estimation of
such high-dimensional objects is rarely feasible in practice, especially
with sample sizes typical in applied work. Consequently, a fully nonparametric
implementation is most practical when $K$ is small. Maintaining tractability
with larger $K$ typically requires additional structure, which is
most natural for outcomes where well-established parametric or semiparametric
models exist.

\section{Extensions}

\subsection{Identifying the Quantile Treatment Effect on the Treated}

The quantile treatment effect on the treated (QTT) is useful for understanding
how a treatment shifts the distribution of outcomes at specific quantiles,
complementing the average effects captured by the ATT. \citet{callaway2019quantile}
show that identifying the QTT in the standard DID framework requires
additional distributional assumptions. By contrast, the framework
in Section \ref{sec:Econometric-Framework} identifies the QTT under
the same assumptions used for identifying the ATT, without requiring
extra restrictions.

For quantile $\tau\in(0,1)$ in period $t\in\{1,\ldots,T_{\text{post}}\}$,
the QTT is defined as
\begin{equation}
\text{QTT}_{t}(\tau)\equiv F^{-1}_{Y_{t}(1)|D=1}(\tau)-F^{-1}_{Y_{t}(0)|D=1}(\tau),
\end{equation}
where $F^{-1}_{Y_{t}(d)|D=1}(\tau)$ denotes the $\tau$-quantile
of the conditional distribution of the potential outcome $Y_{it}(d)$
given $D_{i}=1$ for $d\in\{0,1\}$. The distribution $F_{Y_{t}(1)|D=1}$
is directly observed from the treatment group data. The challenge
lies in identifying the counterfactual distribution $F_{Y_{t}(0)|D=1}$.

Given that $Y_{it}(0)$ and $D_{i}$ are mutually independent conditional
on $U_{i}$ under Assumption 1, this counterfactual distribution can
be expressed as:
\begin{equation}
F_{Y_{t}(0)|D=1}(y)=\int_{u\in{\cal U}}F_{Y_{t}(0)|U}(y|u)f_{U|D=1}(u)du.
\end{equation}
As outlined in the proof of Theorem 1, both $F_{Y_{t}(0)|U}$ and
$f_{U|D}$ can be identified under Assumptions 1--3 and either Assumption
4 or 5, enabling computation of $F_{Y_{t}(0)|D=1}(y)$ and its inverse
to obtain $\text{QTT}_{t}(\tau)$ for any $\tau\in(0,1)$.

\subsection{Identifying Treatment Effect Heterogeneity\label{subsec:Identifying-Treatment-Effect}}

Heterogeneity in individual treatment effects, $Y_{it}(1)-Y_{it}(0)$,
is of key interest in the program evaluation literature. However,
the ATT captures only the mean of these effects, and the QTT does
not reveal their distribution without strong assumptions such as rank
invariance. The baseline framework in Section \ref{sec:Econometric-Framework}
can be extended to identify the conditional average treatment effect
$E[Y_{it}(1)-Y_{it}(0)|U_{i}]$, which captures heterogeneity in treatment
effects across values of $U_{i}$, under modified and additional assumptions.
\begin{assumption}[Conditional Independence, Extended]

$\boldsymbol{Y}^{\text{pre}}_{i}(0)$, $\left(Y_{i0}(0),D_{i}\right)$,
and $\left(\boldsymbol{Y}^{\text{post}}_{i}(0),\right.\allowbreak\left.\boldsymbol{Y}^{\text{post}}_{i}(1)\right)$
are mutually independent conditional on $U_{i}$.
\end{assumption}
Assumption 6 replaces Assumption 1. The key modification is that post-treatment
potential outcomes now appear as a pair $\left(\boldsymbol{Y}^{\text{post}}_{i}(0),\boldsymbol{Y}^{\text{post}}_{i}(1)\right)$
in the conditional independence statement. This allows for arbitrary
correlation between treated and untreated outcomes, which in turn
permits arbitrary distributions of individual treatment effects conditional
on $U_{i}$. However, their dependence with pre-treatment outcomes
and treatment assignment must operate solely through $U_{i}$.
\begin{assumption}[Bounded Densities, Extended]

The joint, marginal, and conditional densities of $(\boldsymbol{Y}^{\text{post}}_{i}(1),U_{i})$
conditional on $D_{i}=1$ are all bounded.
\end{assumption}
Assumption 7 applies the boundedness condition in Assumption 3--(a)
to treated outcomes $\boldsymbol{Y}^{\text{post}}_{i}(1)$. However,
$f_{\boldsymbol{Y}^{\text{post}}(1)|U}$ does not have to satisfy
a completeness condition, unlike $f_{\boldsymbol{Y}^{\text{pre}}(0)|U}$
in Assumption 3--(b) or $f_{\boldsymbol{Y}^{\text{post}}(0)|U}$
in Assumption 4--(a).
\begin{thm}
Under Assumptions 2, 3, 6, 7, and either Assumption 4 or 5, the conditional
average treatment effects $E[Y_{it}(1)-Y_{it}(0)|U_{i}]$ for $t\in\{1,\ldots,T_{\text{post}}\}$
are identified from the joint distribution of $\left(D_{i},Y_{i,-T_{\text{pre}}},\right.\allowbreak\left.\ldots,Y_{i,T_{\text{post}}}\right)$.
\end{thm}
The proof of this theorem requires only one additional step beyond
the baseline framework: identifying the distribution of $\boldsymbol{Y}^{\text{post}}_{i}(1)|U_{i}$.
Assumption 6 implies
\begin{equation}
f_{\boldsymbol{Y}^{\text{pre}},\boldsymbol{Y}^{\text{post}}|D=1}(y,z)=\int_{u\in{\cal U}}f_{\boldsymbol{Y}^{\text{pre}}(0)|U}(y|u)f_{\boldsymbol{Y}^{\text{post}}(1)|U}(z|u)f_{U|D=1}(u)du,\label{eq:integral_eq_ex}
\end{equation}
where $f_{\boldsymbol{Y}^{\text{pre}},\boldsymbol{Y}^{\text{post}}|D=1}$
is known from the data and $f_{\boldsymbol{Y}^{\text{pre}}(0)|U}$
and $f_{U|D=1}$ have already been identified within the baseline
framework. Since Assumption 3--(b) ensures the completeness of $f_{\boldsymbol{Y}^{\text{pre}}(0)|U}$
and Assumption 7 implies $f_{\boldsymbol{Y}^{\text{post}}(1)|D=1}(z)=\int_{u\in{\cal U}}f_{\boldsymbol{Y}^{\text{post}}(1)|U}(z|u)f_{U|D=1}(u)du$
is bounded, the integral equation (\ref{eq:integral_eq_ex}) uniquely
determines $f_{\boldsymbol{Y}^{\text{post}}(1)|U}(z|u)f_{U|D=1}(u)$.
Finally, Assumption 2 ensures that $f_{U|D=1}(u)>0$ for any $u\in{\cal U}$,
enabling the recovery of $f_{\boldsymbol{Y}^{\text{post}}(1)|U}(z|u)$
through division. With the distribution of $\boldsymbol{Y}^{\text{post}}_{i}(1)|U_{i}$
being identified alongside that of $\boldsymbol{Y}^{\text{post}}_{i}(0)|U_{i}$,
$E[Y_{it}(1)-Y_{it}(0)|U_{i}]$ is identified for $t\in\{1,\ldots,T_{\text{post}}\}$.

The identification of $E[Y_{it}(1)-Y_{it}(0)|U_{i}]$ has several
important implications for understanding treatment effects. First,
it enables the recovery of the population average treatment effect
(ATE) through:
\[
E[Y_{it}(1)-Y_{it}(0)]=\int_{u\in{\cal U}}E[Y_{it}(1)-Y_{it}(0)|U_{i}=u]f_{U}(u)du.
\]
Second, it allows identification of the average treatment effect on
the untreated (ATU):
\[
E[Y_{it}(1)-Y_{it}(0)|D_{i}=0]=\int_{u\in{\cal U}}E[Y_{it}(1)-Y_{it}(0)|U_{i}=u]f_{U|D=0}(u)du.
\]
Finally, it provides a lower bound for the variance of individual
treatment effects. Since the inequality
\[
\mathrm{Var}\left(Y_{it}(1)-Y_{it}(0)\right)\ge\mathrm{Var}\left(E[Y_{it}(1)-Y_{it}(0)|U_{i}]\right)
\]
holds by the law of total variance, the conditional average treatment
effect function provides insight into the minimum degree of treatment
effect heterogeneity present in the population.

\subsection{Staggered Treatment Adoption}

The framework from Section \ref{sec:Econometric-Framework} can be
applied to staggered adoption settings to identify group-time specific
average treatment effects. The key is to appropriately define treatment
and control groups for each cohort of interest.

Consider identifying the ATT for a cohort first treated in period
$g+1$ over post-treatment periods $g+1$ through $g+h$. The identification
strategy from Section \ref{sec:Econometric-Framework} applies by
defining this cohort as the treatment group ($D_{i}=1$) and units
never treated or beginning treatment later than period $g+h$ as the
control group ($D_{i}=0$). Units treated before period $g+1$ or
beginning treatment between periods $g+2$ and $g+h$ must be excluded
from the analysis.

Under this construction, suppose that Assumptions 1--4 from Section
\ref{subsec:Setup-and-Assumptions} apply to the included units, where
periods up to $g-1$ serve as pre-treatment periods, period $g$ as
the reference period, and periods $g+1$ through $g+h$ as post-treatment
periods. Theorem 1 then establishes identification of the ATT parameters
for this cohort in periods $g+1$ through $g+h$. Once group-time
effects are identified for various cohorts and periods, they can be
aggregated according to the researcher's objectives (\citealp{callaway2021difference}).

An important limitation arises when units must be excluded and treatment
timing depends on time-varying unobservables. In the hidden Markov
model from Section \ref{subsec:Examples}, if treatment assignment
depends on persistent shocks $\eta_{it}$ that evolve over time, Assumption
1 fails because the exclusion creates a selection problem. Even if
initial treatment at $g+1$ depends only on $\eta_{ig}$, being in
the control group (not treated through period $g+h$) depends on $\eta_{it}$
for $t\in\{g+1,\ldots,g+h-1\}$, creating dependence between outcomes
and treatment status that cannot be conditioned away using only $U_{i}=\eta_{ig}$.
The framework therefore applies most naturally when either (a) no
other units are first treated during periods $g+1$ through $g+h$,
or (b) treatment timing is determined by time-invariant unobservables.

\section{Empirical Illustration\label{sec:Empirical-Illustration}}

This section estimates the impact of job displacement on earnings.
Section \ref{subsec:Standard-Approach-as} employs a standard DID
approach to serve as a benchmark. Section \ref{subsec:Estimation-Without-Parallel}
builds on the framework developed in Section \ref{sec:Econometric-Framework}
to estimate the impact without assuming parallel trends, addressing
potential biases inherent in the standard approach. I use the Sample
of Integrated Labour Market Biographies (SIAB) for this analysis.
The SIAB is a 2\% random sample of all individuals who have ever been
registered in the German social security system, providing detailed
administrative data on earnings and employment histories. It enables
tracking earnings dynamics before and after job displacement over
an extended period, and its large sample size supports the flexible
semiparametric specification employed in Section \ref{subsec:Estimation-Without-Parallel}.

\subsection{Standard Approach as a Benchmark\label{subsec:Standard-Approach-as}}

\subsubsection{Estimation Strategy}

Since the impact of displacement in a particular calendar year is
not of special interest, it is natural to estimate the impact aggregated
across multiple displacement years. In this aggregation, I employ
a stacked estimation approach similar to \citet{jarosch2023searching},
pooling across multiple data sets that correspond to different displacement
years. A data set for displacement year $c$ includes $Y_{i(j,c),t}$,
worker $j$'s earnings in period $t$, where $t=0$ corresponds to
the observation right before year $c$. It also includes $D_{i(j,c)}$,
an indicator for worker $j$'s displacement in year $c$, and $X_{i(j,c)}$,
covariates of worker $j$ observed before year $c$. Pooling these
data sets across different displacement years, I consider a combination
of worker $j$ and displacement year $c$ as individual unit $i(j,c)$
in the pooled data set. 

Taking DID conditional on covariates captures the statistical parameter
\begin{multline}
\Delta^{\text{DID}}_{t}(x)\equiv E\left[Y_{i(j,c),t}-Y_{i(j,c),0}|D_{i(j,c)}=1,X_{i(j,c)}=x\right]\\
-E\left[Y_{i(j,c),t}-Y_{i(j,c),0}|D_{i(j,c)}=0,X_{i(j,c)}=x\right].
\end{multline}
Under a standard conditional parallel trends assumption \citep{heckman1997matching,heckman1998matching,abadie2005semiparametric},
each $\Delta^{\text{DID}}_{t}(x)$ identifies the conditional average
treatment effect on the treated in period $t$. 

As long as covariates $X_{i(j,c)}$ include indicators for displacement
years, each $\Delta^{\text{DID}}_{t}(x)$ represents the impact of
displacement in a particular calendar year conditional on worker observables.
Aggregating $\Delta^{\text{DID}}_{t}(x)$ across covariates, the matching
DID estimand
\begin{equation}
\theta^{\text{DID,M}}_{t}\equiv\int\Delta^{\text{DID}}_{t}(x)dF_{X|D=1}(x)
\end{equation}
identifies the ATT, averaged across different displacement years and
worker observables. I estimate the parameter using a doubly robust
method developed by \citet{sant2020doubly}.

\subsubsection{Sample Construction}

I use displacement years $c\in\{2000,\ldots,2004\}$, with all monetary
values expressed in year-2000 Euros. The sample is restricted to male
workers aged 30--39 in West Germany with vocational training but
no university education, who at the end of year $c-1$ are full-time
employed in jobs subject to social security contributions and have
at least three years of tenure, and who have nonzero, nonmissing earnings
in each of years $c-9$ through $c-1$. All inclusion criteria are
defined as of year $c-1$ and do not depend on future displacement
events. The final sample consists of 49,872 unique individuals ($j$)
and 141,100 individual--displacement-year combinations ($j,c$).
Appendix \ref{subsec:Sample-Construction-Details} discusses the rationale
for these restrictions, the choice of displacement-year window, and
the SIAB processing steps including top-coding. 

An individual $j$ in year $c$ is considered treated ($D_{i(j,c)}=1$)
if his employment spell ends in that year and a claim for unemployment
insurance is filed within 90 days, indicating job displacement due
to involuntary reasons such as layoff or plant closure. This definition
is similar to that used in \citet{jarosch2023searching}. The treatment
group accounts for 2.54\% of all observations. The same worker $j$
may enter the sample as a treated unit for one $c$ and as a control
unit for another, and all reported standard errors account for this
overlap by clustering at the individual worker level. Average earnings
in the year prior to displacement are 28,316 Euros for the treatment
group and 34,802 Euros for the control group. Covariates $X_{i(j,c)}$
include indicators for displacement years and ages in all specifications.
Additional covariates vary across specifications and include region
and occupation indicators, tenure and tenure squared, or pre-treatment
earnings.

\subsubsection{Estimation Results}

\begin{figure}[tph]
\caption{The Estimated Earnings Losses from Displacement (Standard DID)}

\label{Fig:DID}

\vspace{2.0em}
\centering
\begin{threeparttable}

\begin{minipage}[t]{0.49\columnwidth}%
\centering (A) Baseline Specification

\includegraphics[width=1\columnwidth]{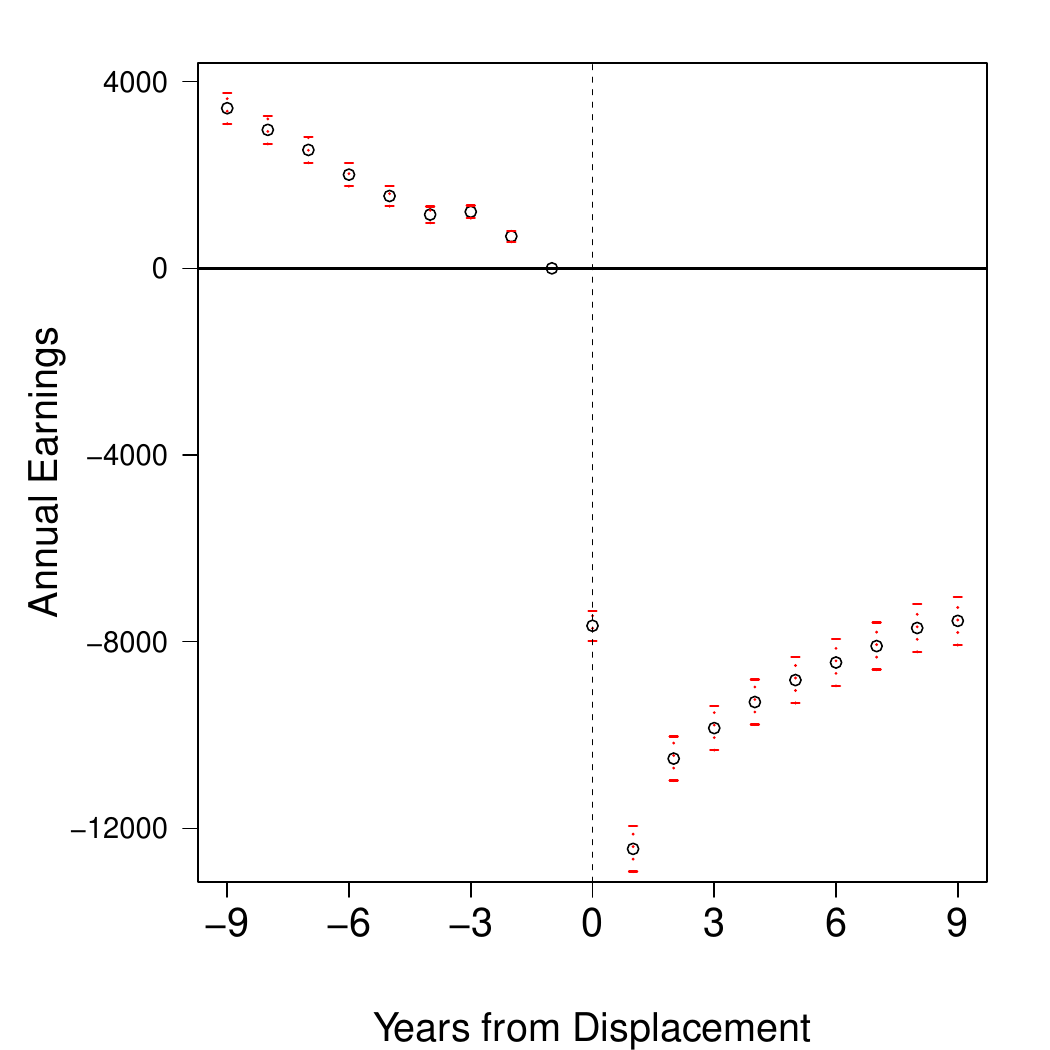}%
\end{minipage}%
\begin{minipage}[t]{0.49\columnwidth}%
\centering (B) With Additional Controls

\includegraphics[width=1\columnwidth]{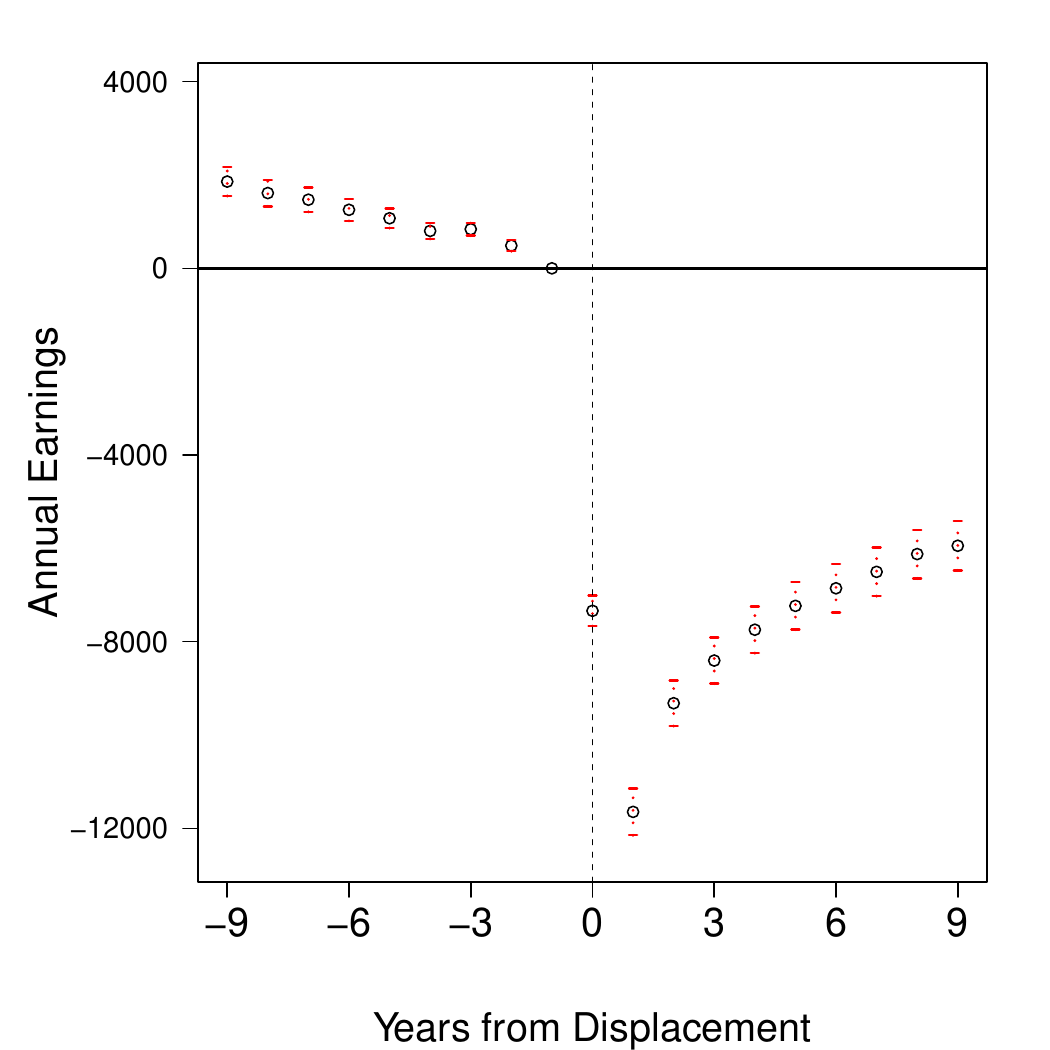}%
\end{minipage}

\vspace{3.0em}

\begin{minipage}[t]{0.49\columnwidth}%
\centering (C) Pre-Treatment Earnings as Covariates

\includegraphics[width=1\columnwidth]{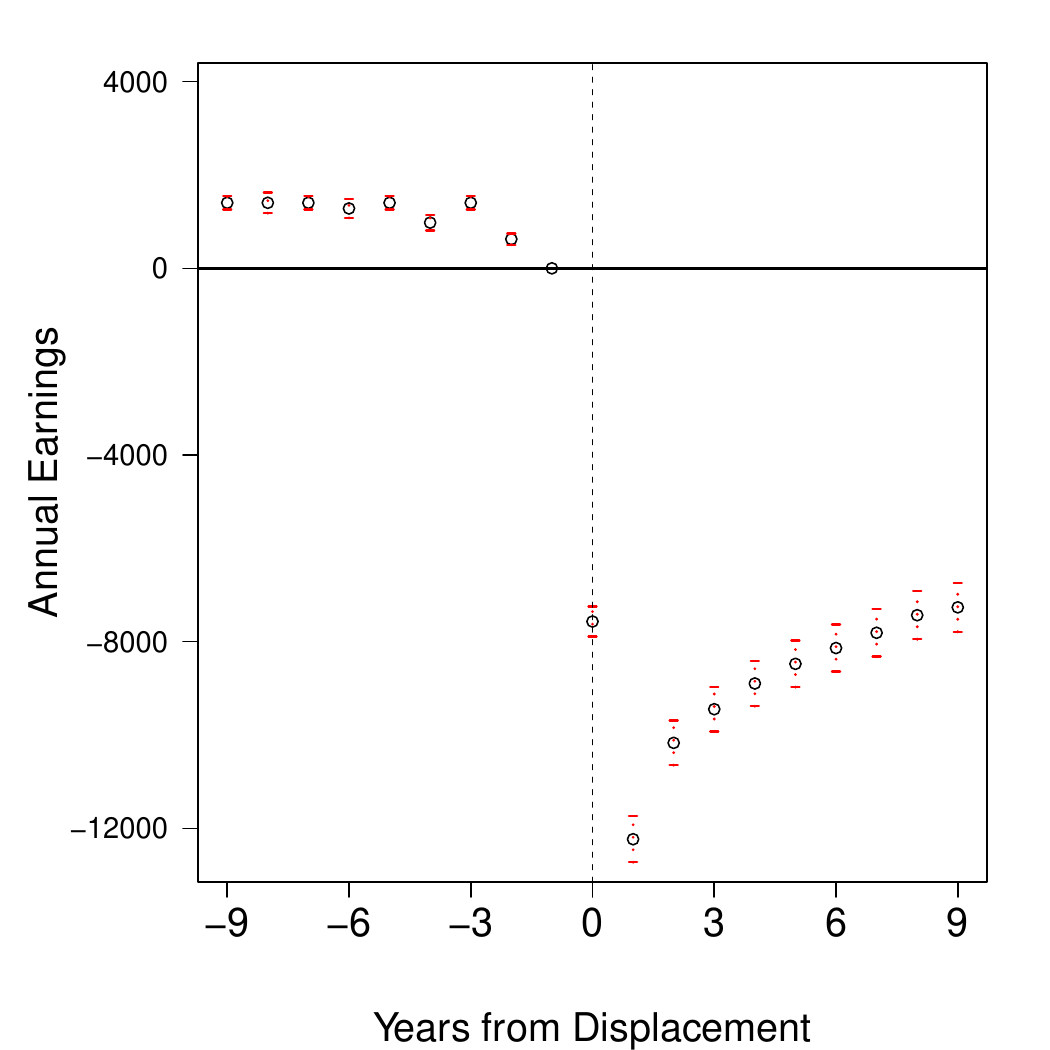}%
\end{minipage}%
\begin{minipage}[t]{0.49\columnwidth}%
\centering (D) Unit-Specific Linear Trends

\includegraphics[width=1\columnwidth]{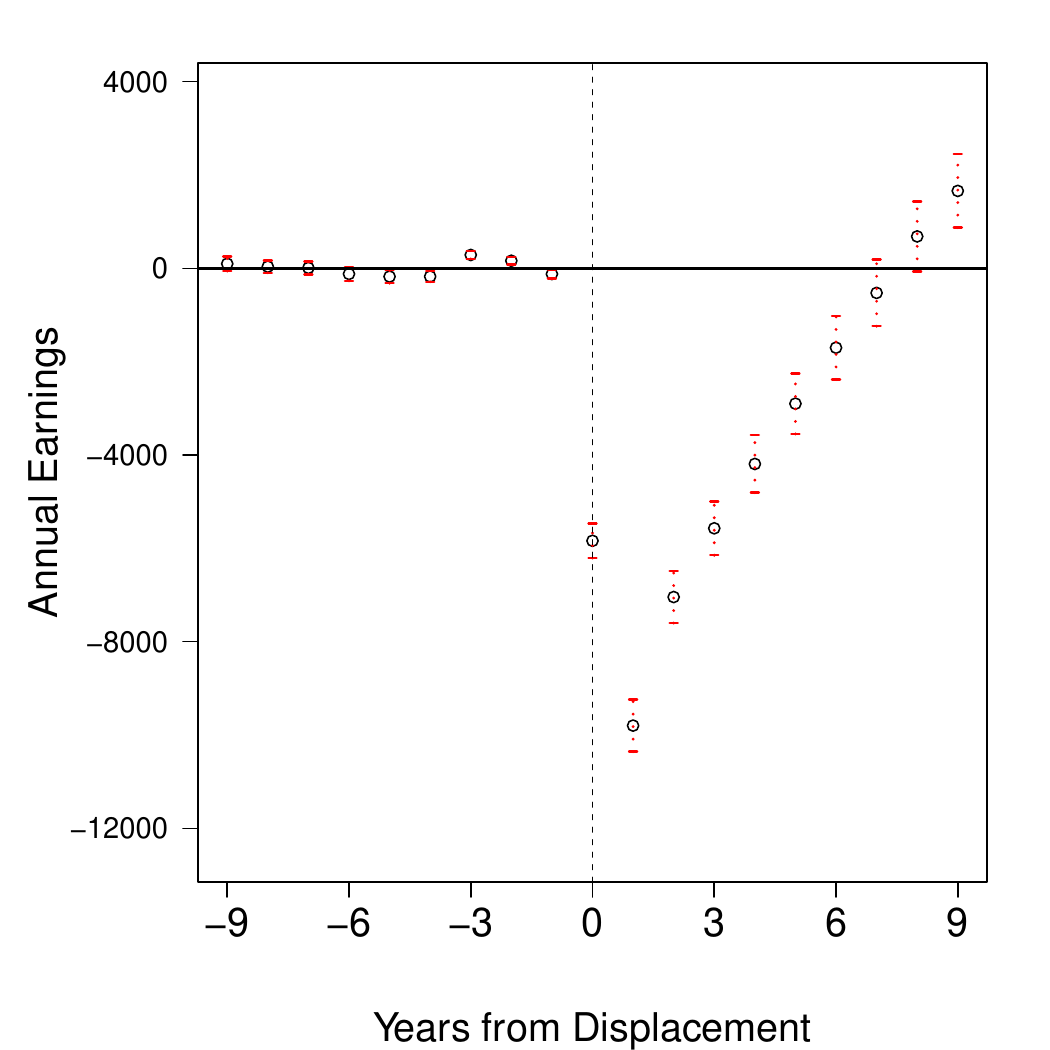}%
\end{minipage}

\vspace{1.0em}
\begin{tablenotes}
\footnotesize

\item Note: Estimates are from doubly robust matching DID (Panels
A--C; \citealp{sant2020doubly}) or a regression with unit-specific
trends (Panel D; equation (\ref{eq:unit-specific-trends})). All panels
include age and calendar year indicators as covariates. Panel B additionally
includes region, occupation, and tenure controls, and Panel C additionally
includes lagged earnings. Vertical lines denote 95\% confidence intervals
with standard errors clustered at the individual worker level. Earnings
are expressed in year-2000 Euros. Exact numerical values underlying
the graphical results are presented in Appendix \ref{subsec:Point-Estimates-for}.

\end{tablenotes}
\end{threeparttable}
\end{figure}

Figure \ref{Fig:DID} presents the estimated effects of job displacement
on earnings. Panels A--C plot the matching DID estimates $\hat{\theta}^{\text{DID,M}}_{t}$
using different sets of covariates. Panel D presents estimates from
a regression with unit-specific linear trends, described below. Since
the reference period $t=0$ corresponds to a year before displacement,
estimates for pre-treatment periods ($t<0$) appear at event times
$-2$, $-3$, ..., $-9$, and those for post-treatment periods ($t>0$)
appear at event times 0, 1, ..., 9. Exact point estimates and standard
errors are reported in Appendix \ref{subsec:Point-Estimates-for}.

Panel A presents estimates from the baseline specification, which
includes indicators for age and calendar year at $t=0$ as covariates.
The estimate for one year after displacement suggests earnings losses
of 12,440 Euros, which amounts to 44\% of the pre-displacement average
earnings. Long-run estimates indicate persistent earnings losses,
with the estimated effect remaining at 7,554 Euros nine years after
displacement. However, the pre-treatment estimates show a clear downward
trend, with earnings in the treatment group declining relative to
the control group in the years preceding displacement. This pre-trend
calls into question the validity of the parallel trends assumption.
Such pre-trends are not unusual in the displacement literature; for
example, \citet{jacobson1993earnings} document pre-trends amounting
to roughly one-sixth of average earnings over the five years preceding
displacement.

Panel B adds indicators for region and occupation, as well as tenure
and its square, to the baseline covariates. The estimated loss one
year after displacement declines modestly to 11,647 Euros and the
effect nine years after to 5,945 Euros. While the pre-trend also persists,
interpreting the pre-trend diagnostics in this specification is complicated
by the fact that the added covariates are measured at $t=0$ and may
be jointly determined with pre-trends, creating a potential ``bad
control'' problem.

Panel C adds pre-treatment outcomes, earnings from three, five, seven,
and nine years before the displacement year, to the baseline specification.
While this approach is often employed in the job displacement and
job training evaluation literature, lagged outcomes inherently contain
transitory shocks in addition to persistent heterogeneity, limiting
their effectiveness as covariates. The estimated loss one year after
displacement is 12,233 Euros and the effect nine years after is 7,265
Euros, marginally lower than Panel A. Because lagged outcomes mechanically
correlate with pre-trends, the pre-trend diagnostics in this specification
cannot be meaningfully interpreted; they are shown only for completeness.

Panel D reports estimates from a regression with unit-specific linear
trends,
\begin{equation}
Y_{it}=\alpha_{i}+\sum^{9}_{k=-9}\beta_{k}d^{(k)}_{it}+X_{i}'\gamma_{t}+\delta_{i}t+\varepsilon_{it},\label{eq:unit-specific-trends}
\end{equation}
where $d^{(k)}_{it}=1$ if unit $i$ in period $t$ is $k$ years
from displacement and $X_{i}$ includes the baseline covariates (age
and displacement year indicators). When unit-specific trends $\delta_{i}t$
are included, any linear pattern in the event-time coefficients $\{\beta_{k}\}$
cannot be separately identified from these trends. As a result, at
least two independent restrictions on the pre-treatment coefficients
are required, rather than the usual reference-period normalization
$\beta_{-1}=0$. The particular restrictions chosen affect the estimated
post-treatment coefficients (\citealp{miller2023introductory}), so
the choice must be stated and justified. A central premise of the
unit-specific trend specification is that linear trends fully capture
preexisting differences between treated and control units, leaving
no residual pre-trends. Under this premise, one would impose $\beta_{k}=0$
for all $k<0$. To retain information on nonlinear pre-trends, Panel
D instead imposes two weaker constraints: a zero-average condition,
$\sum^{-1}_{k=-9}\beta_{k}=0$, and a flat-trend condition, $\sum^{-1}_{k=-9}\beta_{k}(k+5)=0$.
These restrictions yield exactly the same post-treatment coefficients
$\{\beta_{k}\}_{k\ge0}$ as the stronger $\beta_{k}=0$ for all $k<0$
restriction.\footnote{The equivalence arises because the average and trend of $\{\beta_{k}\}_{k<0}$
affect the estimates of unit fixed effects $\alpha_{i}$ and unit-specific
trends $\delta_{i}$, which in turn influence the post-treatment coefficients
$\{\beta_{k}\}_{k\ge0}$. The stronger $\beta_{k}=0$ for all $k<0$
restriction also imposes zero average and zero trend on the pre-treatment
coefficients, and thus has an identical impact on $\{\beta_{k}\}_{k\ge0}$.}

The results in Panel D are striking: the estimated effects reverse
sign in later years, suggesting earnings gains eight or nine years
after displacement. This pattern is difficult to reconcile with economic
theory or prior evidence and likely reflects misspecification. Alternative
normalizations such as $\beta_{-9}=\beta_{-1}=0$ produce similar
results, suggesting that the issue stems from the unit-specific trend
specification rather than the chosen normalization. 

\subsection{Estimation Without Parallel Trends\label{subsec:Estimation-Without-Parallel}}

I now consider an alternative estimation approach that does not rely
on the parallel trends assumption, building on the identification
framework developed in Section \ref{sec:Econometric-Framework}.

\subsubsection{Estimation Strategy}

I specify a semiparametric model with two-dimensional unobserved heterogeneity,
$U_{i}=(U_{1i},U_{2i})$, where each dimension is independently drawn
from a uniform distribution on $(0,1)$. This distributional specification
serves as a normalization rather than a substantive assumption; as
shown in Appendix \ref{subsec:Normalization-of-Unobserved}, regardless
of the true joint distribution of $U_{i}$, the model can be equivalently
represented with independent uniform $U_{i}$ through appropriate
transformations.

The choice of two dimensions balances richness of heterogeneity and
computational feasibility: one dimension fails to capture the differential
earnings growth in Panel A of Figure \ref{Fig:DID}, while $K>2$
is less practical in this semiparametric setup.\footnote{The specification developed below already involves 189 parameters
at $K=2$.} Since violations of parallel trends arise when the two groups differ
in the growth of counterfactual earnings rather than in levels alone,
two dimensions is the minimal choice that accommodates the heterogeneity
at issue, allowing earnings growth to vary across workers independently
of earnings levels. The unit-specific linear trend specification of
Panel D of Figure \ref{Fig:DID} serves the same purpose in a much
more restrictive way; the two-dimensional specification nests it as
a special case while permitting arbitrary nonlinear and nonadditive
dependence of outcomes on $U_{i}$. As a robustness exercise, Appendix
\ref{Asec:Three-Dimensional-Specification} reports results for a
specification with $K=3$.

The identification results in Section \ref{sec:Econometric-Framework}
require $K\le\min\{T_{\text{pre}},T_{\text{post}}\}$, which is satisfied
with $K=2$ as long as at least two pre-treatment and two post-treatment
periods are included in the data. While using larger $T_{\text{pre}}$
and $T_{\text{post}}$ would in principle yield greater precision,
semiparametric modeling of conditional densities for high-dimensional
outcome vectors becomes increasingly intractable. I therefore set
$T_{\text{pre}}=T_{\text{post}}=2$, using two pre-treatment observations
and two post-treatment observations.

As described in Section \ref{subsec:Standard-Approach-as}, the sample
provides eight years of pre-treatment earnings (two to nine years
before displacement) and ten years of post-treatment earnings (zero
to nine years after displacement). Selecting which observations to
include involves balancing two considerations. First, Assumption 1
is more credible when the chosen observations are temporally separated
from the reference period (one year before displacement), reducing
the likelihood that serial correlation in untreated outcomes arises
through channels beyond what is captured by $U_{i}$. Second, observations
that are too close together may provide less independent information
about distinct dimensions of $U_{i}$, reducing precision.

Balancing these considerations, I use earnings from five and nine
years before displacement as the two pre-treatment outcomes, and earnings
from five and nine years after displacement as the two post-treatment
outcomes. The selected observations are at least four years apart
from the reference period and each other within each window.

I estimate the model of untreated outcomes $\left\{ Y_{it}(0)\right\} ^{T_{\text{post}}}_{t=-T_{\text{pre}}}$
and treatment assignment $D_{i}$ given $(X_{i},U_{i})$ using a maximum
likelihood method suggested in Section \ref{subsec:Estimation}.\footnote{Since the same worker enters as a unit for multiple displacement years,
the objective function is a pseudo-likelihood; the estimator remains
consistent and standard errors are clustered at the individual worker
level.} Then, I recover the ATT parameters using equation (\ref{eq:ATT_equation})
with a modification. While Section \ref{sec:Econometric-Framework}
has kept covariates $X_{i}$ implicit, they must now be made explicit.
With covariates, equation (\ref{eq:ATT_equation}) should be modified
as:
\begin{equation}
\theta^{\text{ATT}}_{t}=\theta^{\text{M}}_{t}-\int\int_{u\in{\cal U}}E\left[Y_{it}(0)|U_{i}=u,X_{i}=x\right]\left(f_{U|D=1,X}(u|x)-f_{U|D=0,X}(u|x)\right)dudF_{X|D=1}(x),\label{eq:ATT_x}
\end{equation}
where 
\begin{equation}
\theta^{\text{M}}_{t}\equiv\int\left\{ E\left[Y_{it}|D_{i}=1,X_{i}=x\right]-E\left[Y_{it}|D_{i}=0,X_{i}=x\right]\right\} dF_{X|D=1}(x)
\end{equation}
is the matching ATT estimand in a cross-sectional treatment effect
framework, which can be estimated from the data using a doubly robust
method. Given that I follow \citet{sant2020doubly} for standard DID
estimates, I use their method with outcome changes replaced by outcome
levels to estimate $\theta^{\text{M}}_{t}$. The components $E\left[Y_{it}(0)|U_{i}=u,X_{i}=x\right]$
and $f_{U|D=d,X}(u|x)$ are taken from the estimated model, while
$F_{X|D=1}(x)$ is taken from the empirical distribution in the data.

Covariates $X_{i}$ include age and calendar year indicators, following
the baseline specification in Section \ref{subsec:Standard-Approach-as}.
In estimating $\theta^{\text{M}}_{t}$, both covariates are used to
ensure flexible matching. In the model specification below, calendar
year effects are omitted and age effects enter quadratically to limit
the number of parameters.

\subsubsection{Empirical Specification}

I now specify a semiparametric model for the displacement probability
and conditional earnings densities. The probability of displacement
is specified using a partially linear model:
\begin{equation}
\Pr\left[D_{i}=1|U_{i},X_{i}\right]=F_{\text{logit}}\left(\sum_{0\le j+k\le3}\alpha^{d}_{jk}T_{j}(U_{1i})T_{k}(U_{2i})+X_{i}'\beta^{d}\right),\label{eq:prob_treated}
\end{equation}
where $F_{\text{logit}}(s)=\frac{\exp(s)}{1+\exp(s)}$, and $T_{j}(u)$
is the $j$-th order Chebyshev polynomial on $(0,1)$.

The conditional density of pre-treatment outcomes is specified as:
\begin{equation}
f_{\ln Y_{-1}(0),\ln Y_{-2}(0)|U,X}\left(y_{-1},y_{-2}|u,x\right)=\frac{h_{2}\left(\frac{y_{-1}-\mu_{-1}(u,x)}{\sigma_{-1}(u,x)},\frac{y_{-2}-\mu_{-2}(u,x)}{\sigma_{-2}(u,x)};\boldsymbol{\omega}^{\text{pre}}\right)}{\sigma_{-1}(u,x)\sigma_{-2}(u,x)},
\end{equation}
where $y_{-1}$ and $y_{-2}$ represent log earnings from five and
nine years before displacement ($t=-1$ and $t=-2$), respectively.
The function $h_{2}$ is a two-dimensional density modeled using a
sieve approximation: 
\begin{equation}
h_{2}\left(v_{1},v_{2};\boldsymbol{\omega}\right)=\frac{\phi(v_{1})\phi(v_{2})\left(1+\sum_{1\le j+k\le4}\omega_{jk}\frac{H_{j}(v_{1})}{\sqrt{j!}}\frac{H_{k}(v_{2})}{\sqrt{k!}}\right)^{2}}{1+\sum_{1\le j+k\le4}\omega^{2}_{jk}},\label{eq:density_hermite}
\end{equation}
where $\phi(v)$ denotes the standard normal density and $H_{j}(v)$
is the $j$-th order Hermite polynomial.\footnote{The Hermite sieve order exceeds the one for Chebyshev by one. Chebyshev
terms enter many partially linear functions, and raising their order
to 4 would add 65 parameters to the model.} The denominator in equation (\ref{eq:density_hermite}) ensures that
the density integrates to one.

This specification achieves tractability by letting location $\mu_{t}(u,x)$
and scale $\sigma_{t}(u,x)$ vary across $(u,x)$, while keeping higher-order
features captured by the parameter $\boldsymbol{\omega}^{\text{pre}}$
constant. I specify the functions $\mu_{t}(u,x)$ and $\ln\sigma_{t}(u,x)$
to be partially linear in $(u,x)$, mirroring equation (\ref{eq:prob_treated}):
\begin{eqnarray}
\mu_{t}(u,x) & = & \sum_{0\le j+k\le3}\alpha^{\mu}_{jkt}T_{j}(u_{1})T_{k}(u_{2})+x'\beta^{\mu}_{t},\label{eq:mu_t}\\
\ln\sigma_{t}(u,x) & = & \sum_{0\le j+k\le3}\alpha^{\sigma}_{jkt}T_{j}(u_{1})T_{k}(u_{2})+x'\beta^{\sigma}_{t}.\label{eq:sigma_t}
\end{eqnarray}

The conditional density of the reference period outcome is specified
as:
\begin{eqnarray}
f_{\ln Y_{0}(0)|U,X,D}\left(y|u,x,d\right) & = & \frac{h_{1}\left(\frac{y-\mu_{0}(u,x,d)}{\sigma_{0}(u,x,d)};\boldsymbol{\omega}^{0}\right)}{\sigma_{0}(u,x,d)},\\
h_{1}(v;\boldsymbol{\omega}) & = & \frac{\phi(v)\left(1+\sum^{4}_{j=1}\omega_{j}\frac{H_{j}(v)}{\sqrt{j!}}\right)^{2}}{1+\sum^{4}_{j=1}\omega^{2}_{j}}.
\end{eqnarray}
Since the reference period outcome is one-dimensional, the sieve approximation
$h_{1}$ uses only univariate Hermite polynomials. The location $\mu_{0}(u,x,d)$
and scale $\sigma_{0}(u,x,d)$ functions depend on treatment status
$d$ in addition to $(u,x)$, consistent with Assumption 1. I specify
the functions $\mu_{0}(u,x,d)$ and $\ln\sigma_{0}(u,x,d)$ to be
nonlinear in $u$ via Chebyshev polynomials and linear in $(x,d)$,
analogous to equations (\ref{eq:mu_t}) and (\ref{eq:sigma_t}). The
density depends on $u$ through both $\mu_{0}$ and $\sigma_{0}$
rather than a single index, consistent with Assumption 4--(c).

The conditional density of post-treatment untreated outcomes requires
a different specification than that for pre-treatment outcomes. While
a positive-earnings restriction is applied to the sample for $t\le0$,
this restriction does not apply to $t>0$, as having no earnings is
a natural outcome following displacement. I model post-treatment untreated
outcomes as 
\begin{equation}
Y_{it}(0)=(1-Z_{it})\exp\left(\tilde{Y}_{it}\right),
\end{equation}
where $Z_{it}$ is an indicator for zero earnings in period $t$ and
$\tilde{Y}_{it}$ denotes log earnings.

The conditional probability $\Pr[Z_{i1}=1|U_{i},X_{i}]$ follows the
partially linear specification in equation (\ref{eq:prob_treated}).
The conditional probability $\Pr[Z_{i2}=1|U_{i},X_{i},Z_{i1}]$ is
specified analogously, with $U_{i}$ entering nonlinearly via Chebyshev
polynomials and $(X_{i},Z_{i1})$ entering linearly.

The conditional density of log earnings is given by:
\begin{equation}
f_{\tilde{Y}_{1},\tilde{Y}_{2}|U,X,Z_{1},Z_{2}}\left(y_{1},y_{2}|u,x,z_{1},z_{2}\right)=\frac{h_{2}\left(\frac{y_{1}-\mu_{1}(u,x,z_{2})}{\sigma_{1}(u,x,z_{2})},\frac{y_{2}-\mu_{2}(u,x,z_{1})}{\sigma_{2}(u,x,z_{1})};\boldsymbol{\omega}^{\text{post}}\right)}{\sigma_{1}(u,x,z_{2})\sigma_{2}(u,x,z_{1})},
\end{equation}
where $h_{2}$ is defined in equation (\ref{eq:density_hermite}).
The location and scale functions are specified to be nonlinear in
$u$ and linear in $(x,z)$, analogous to equations (\ref{eq:mu_t})
and (\ref{eq:sigma_t}). Since $\tilde{Y}_{it}$ affects the actual
outcome $Y_{it}(0)$ only when $Z_{it}=0$, the specification allows
the location and scale to be shifted by the zero-earnings indicator
in the other period only.

\subsubsection{Estimation Results}

Table \ref{Table:ATT_alt} presents the estimated effects of job displacement
on earnings five and nine years after displacement. The first column
reproduces the matching DID estimates from the baseline specification
in Section \ref{subsec:Standard-Approach-as}. The second column reports
the ATT estimates obtained using the alternative approach that does
not rely on parallel trends. The third column shows the difference
between the two sets of estimates.

\begin{table}

\caption{The Estimated Earnings Losses from Displacement (Standard vs. Alternative)}

\label{Table:ATT_alt}

\centering
\begin{threeparttable}
\begin{centering}
\begin{tabular}{ccccc}
 &  &  &  & \tabularnewline
\hline 
Years &  & Standard & Alternative: & Difference:\tabularnewline
from &  & DID: $\hat{\theta}^{\text{DID,M}}_{t}$ & $\hat{\theta}^{\text{ATT}}_{t}$ & $\hat{\theta}^{\text{DID,M}}_{t}-\hat{\theta}^{\text{ATT}}_{t}$\tabularnewline
Displacement &  & (1) & (2) & (3)\tabularnewline
\cline{1-1}\cline{3-5}
\multirow{2}{*}{5} &  & --8,823 & --5,851 & --2,972\tabularnewline
 &  & \,\,\,\,\,\,(251) & \,\,\,\,\,\,(525) & \,\,\,\,\,\,(467)\tabularnewline
\multirow{2}{*}{9} &  & --7,554 & --4,221 & --3,334\tabularnewline
 &  & \,\,\,\,\,\,(264) & \,\,\,\,\,\,(570) & \,\,\,\,\,\,(519)\tabularnewline
\hline 
\end{tabular}
\par\end{centering}
\begin{tablenotes}
\footnotesize

\item Note: Standard DID estimates correspond to Panel A of Figure
\ref{Fig:DID}. Standard errors (in parentheses) are clustered at
the individual worker level. Outcomes are measured by annual earnings
in year-2000 Euros.

\end{tablenotes}
\end{threeparttable}

\end{table}

The alternative approach yields considerably smaller estimated earnings
losses than the standard DID method. Five years after displacement,
it estimates a reduction in earnings of 5,851 Euros, 66\% of the standard
DID estimate of 8,823 Euros. Nine years after displacement, the gap
widens further: the estimated loss of 4,221 Euros is 56\% of the 7,554-Euro
reduction suggested by the standard DID estimate.

These results suggest that the standard DID estimates may substantially
overstate the long-run earnings impact of job displacement. The divergence
between the two estimates is consistent with the presence of preexisting
downward trends in the treatment group, as documented in Panel A of
Figure \ref{Fig:DID}. The difference reflects the alternative approach's
ability to account for differential trends across treatment and control
groups. Nevertheless, this approach does not simply extrapolate pre-trends
linearly; as shown in Panel D of Figure \ref{Fig:DID}, such mechanical
extrapolation through unit-specific trends produces implausible results.
Rather, the approach exploits the structure of pre-treatment earnings
dynamics to recover the distribution of unobservables governing selection
and their relationship to untreated outcomes.

A natural concern is that two-dimensional unobserved heterogeneity
may be insufficient to capture the true data-generating process. If
the underlying model involves higher-dimensional unobservables, the
estimated specification recovers a best two-dimensional approximation
rather than the true model. Appendix \ref{Asec:Three-Dimensional-Specification}
addresses this concern directly by estimating a three-dimensional
version of the model. It yields estimated earnings losses of 4,366
Euros five years after displacement and 4,326 Euros nine years after
displacement, against the baseline estimates of 5,851 and 4,221 Euros
in Table \ref{Table:ATT_alt}. Raising the dimension of the unobservable
from two to three therefore leaves the long-run conclusion unchanged.

\subsubsection{Treatment Effects for Additional Years}

Table \ref{Table:ATT_alt} uses post-treatment observations from five
and nine years after displacement, yielding earnings loss estimates
for these two years only. Ideally, one would estimate the ATTs for
all post-treatment years jointly by modeling the conditional density
of the full vector of post-treatment untreated outcomes given $(U_{i},X_{i})$.
As noted earlier, however, semiparametric modeling of conditional
densities for high-dimensional outcome vectors is intractable. To
obtain treatment effect estimates for additional years, I therefore
re-estimate the model using different pairs of post-treatment observations,
maintaining $T_{\text{post}}=2$. Specifically, I use earnings from
years $k$ and 9 for $k\in\{0,1,\ldots,7\}$, which provides estimates
for years 0 through 7 after displacement. For year 8, pairing with
year 9 results in estimation difficulties due to the proximity of
observations, so I instead pair with year 5. The estimate for year
9 comes from pairing years 5 and 9, as in Table \ref{Table:ATT_alt}.

\begin{figure}[tp]
\caption{The Estimated Earnings Losses from Displacement Across Different Years}

\label{Fig:Alt}

\vspace{1.0em}
\centering
\begin{threeparttable}

\includegraphics[width=0.67\columnwidth]{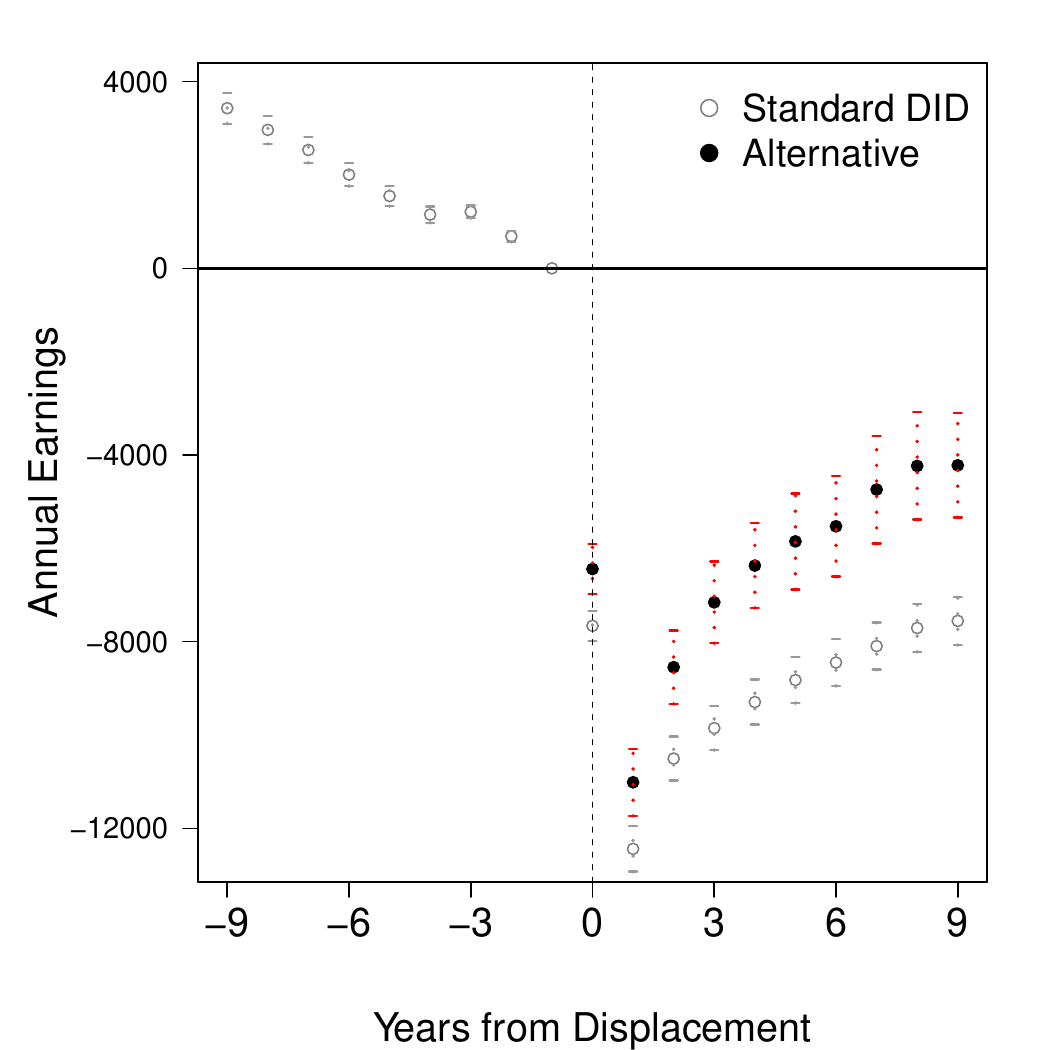}

\vspace{1.0em}
\begin{tablenotes}
\footnotesize

\item Note: Estimates from the alternative method are obtained using
different pairs of post-treatment observations: years $k$ and 9 for
$k\in\{0,1,\ldots,7\}$, years 5 and 8 for year 8, and years 5 and
9 for year 9. Standard DID estimates correspond to Panel A of Figure
\ref{Fig:DID}. Vertical lines denote 95\% confidence intervals with
standard errors clustered at the individual worker level. Earnings
are expressed in year-2000 Euros. Exact numerical values underlying
the graphical results are presented in Appendix \ref{subsec:Point-Estimates-with-Year-Pair}.

\end{tablenotes}
\end{threeparttable}
\end{figure}

Figure \ref{Fig:Alt} plots the estimated earnings losses for each
year after displacement alongside the standard DID estimates from
Panel A of Figure \ref{Fig:DID}. The estimates from the alternative
method are consistently smaller in magnitude than the corresponding
standard DID estimates. The deviation is more pronounced for long-run
effects. Exact point estimates and standard errors are reported in
Appendix \ref{subsec:Point-Estimates-with-Year-Pair}.

A notable feature of Figure \ref{Fig:Alt} is the absence of pre-treatment
estimates for the alternative approach. Instead, pre-treatment observations
are used to identify the distribution of unobserved heterogeneity.
This mirrors the issue in Panel C of Figure \ref{Fig:DID}, where
conditioning on lagged outcomes makes pre-trend diagnostics uninterpretable,
and Panel D, where unit-specific linear trends leave linear pre-trends
unidentified. In all these cases, pre-treatment observations serve
an identification role rather than a diagnostic one.

\section{Conclusion}

This paper develops an identification strategy for treatment effects
without relying on parallel trends. The key insight is that repeated
observations of untreated outcomes serve as multiple noisy measurements
of unobserved heterogeneity, enabling identification through nonclassical
measurement error techniques. This approach accommodates multidimensional
unobservables and arbitrary functional forms, relaxing the implicit
restrictions of the standard DID approach.

The framework has several limitations. First, the dimension of unobserved
heterogeneity cannot exceed the number of pre-treatment observations
or the number of post-treatment observations. Second, it restricts
serial correlation: the reference period outcome must be independent
of other outcomes conditional on unobservables, though correlation
within pre-treatment or post-treatment outcomes is permitted. This
typically requires sufficient temporal separation of the reference
period, increasing data requirements. Third, implementation requires
specifying and estimating a model of the joint distribution of outcomes,
treatment assignment, and unobservables, which is considerably more
complex than standard regression-based DID estimation.

Despite these constraints, the framework offers a viable alternative
when the parallel trends assumption is implausible and researchers
have access to rich panel data with sufficient sample size. By providing
theoretical foundations for leveraging pre-treatment outcomes beyond
simple covariate adjustment, it extends the toolkit for causal inference
when unobserved heterogeneity plays a complex role in shaping outcomes.

\bibliographystyle{econ-econometrica}
\bibliography{NLDID}

\appendix
\setcounter{table}{0}
\setcounter{equation}{0}
\renewcommand{\thetable}{A.\arabic{table}}
\renewcommand{\theequation}{A.\arabic{equation}}

\section{Estimation Details\label{Asec:Estimation-Details}}

\subsection{Sample Construction Details\label{subsec:Sample-Construction-Details}}

In the literature on job displacement, the standard DID approach is
frequently used to estimate the impact of job loss in year $c$ on
earnings in year $c+k$. While the definition of the treatment group
(workers displaced in a specific year $c$) is consistent, definitions
of the control group vary across studies. One approach uses workers
not displaced in year $c$ as the control group (e.g., \citealp{davis2011recessions,jarosch2023searching}).
Another approach uses those who were never displaced between years
$c$ and $c+k$ as the control group, excluding individuals displaced
between years $c+1$ and $c+k$ (e.g., \citealp{jacobson1993earnings,couch2010earnings}).
\citet{krolikowski2018choosing} provides a detailed discussion comparing
the two approaches, highlighting a potential problem of the latter.
I adopt the first approach because it is consistent with the standard
notion of potential outcomes. The most natural definition of the impact
of displacement in year $c$ on earnings in year $c+k$ is the difference
between: (i) earnings that would be realized in year $c+k$ if the
worker were displaced in year $c$, and (ii) earnings that would be
realized in year $c+k$ if the worker were not displaced in year $c$.
By design, both of these counterfactuals must allow for the possibility
of displacement events after year $c$. The second approach, despite
its potential conceptual issue, is motivated by the concern that this
comparison may not be of interest if the control group is so comparable
to the treatment group that they too are likely to be displaced shortly
after. In my data, I do not observe an immediate spike in displacement
hazard in the control group.

The choice of 2000 as the starting year reflects that the SIAB began
to cover marginal part-time employment from April 1999, which is essential
to accurately measure the consequences of job loss. The endpoint 2004
is chosen to allow for a sufficiently long post-treatment period,
and to exclude displacement during or immediately before the Great
Recession, when labor market conditions and adjustment paths were
markedly different. The age, region, and education restrictions reflect
my framework's requirement to keep unobserved heterogeneity within
a small number of dimensions---a condition that could be violated
if a highly diverse set of workers were included. The employment-history
restrictions (full-time and at least three years of tenure at year
$c-1$, and nine years of nonzero, nonmissing earnings) focus the
sample on workers with stable pre-displacement employment histories,
for whom job loss is likely to entail economically meaningful consequences.
The positive-earnings restriction is compatible with Assumption 1.
If the assumption holds in the unselected sample with all years included,
it holds in the selected sample used for estimation; conditional independence
implies that selecting on earnings in some years does not affect the
conditional distribution of earnings in other years given $U_{i}$.

The original SIAB data are spell-based. I use code provided by \citet{dauth2020preparing}
to convert the spell data into a yearly panel spanning nine years
before and after each displacement year, aggregating daily wages within
each year to obtain annual earnings. Following the standard practice
in studies using the SIAB, workers who are not observed in a given
year are assigned zero earnings, reflecting the lack of earnings covered
by social security records. Because the SIAB data exclude civil servants
and self-employed workers, transitions to civil service or self-employment
would appear as a drop to zero in $Y_{i(j,c),t}$. This measurement
issue does not complicate the econometric interpretation as long as
the effect is defined consistently as the impact of displacement on
earnings from employment subject to social security contributions.
The economic interpretation requires caution, however, because the
measured ``loss'' may reflect occupational or sectoral transitions
rather than an actual decline in total earnings.

Daily wages are top-coded based on the social security contribution
ceiling. For workers with a single job, annual earnings are therefore
capped at (maximum daily wage) \texttimes{} (number of days in the
year). However, workers holding multiple jobs simultaneously can exceed
this cap because each job's daily wage is capped separately before
aggregation. To ensure consistent top-coding across single-job and
multiple-job holders, I impose a uniform earnings cap of (maximum
daily wage) \texttimes{} (number of days in the year) for all individuals.
The share of observations at the earnings cap is 1.9\% nine years
before the displacement year, 5.5\% one year before the displacement
year, and 7.7\% nine years after the displacement year.

\subsection{Standard DID Results\label{subsec:Point-Estimates-for}}

Table \ref{Table:ATT_full} presents the exact point estimates and
standard errors corresponding to the graphical results shown in Figure
\ref{Fig:DID} in Section \ref{subsec:Standard-Approach-as}. Columns
A--D correspond to Panels A--D in the figure.

\begin{table}
\caption{The Estimated Earnings Losses from Displacement (Standard DID)}

\label{Table:ATT_full}

\centering
\begin{threeparttable}
\begin{centering}
\begin{tabular}{cccccc}
 &  &  &  &  & \tabularnewline
\hline 
Years from &  & \multicolumn{4}{c}{Specifications}\tabularnewline
\cline{3-6}
displacement &  & (A) & (B) & (C) & (D)\tabularnewline
\cline{1-1}\cline{3-6}
--9 &  & \,\,\,\,\,\,3,431 (169) & \,\,\,\,\,\,1,859 (158) & \,\,\,\,\,\,1,404 \,\,\,(71) & \,\,\,\,\,\,\,\,\,\,\,\,\,\,98 \,\,\,(79)\tabularnewline
--8 &  & \,\,\,\,\,\,2,967 (154) & \,\,\,\,\,\,1,613 (145) & \,\,\,\,\,\,1,404 (113) & \,\,\,\,\,\,\,\,\,\,\,\,\,\,37 \,\,\,(67)\tabularnewline
--7 &  & \,\,\,\,\,\,2,537 (140) & \,\,\,\,\,\,1,470 (133) & \,\,\,\,\,\,1,404 \,\,\,(71) & \,\,\,\,\,\,\,\,\,\,\,\,\,\,\,\,\,9 \,\,\,(71)\tabularnewline
--6 &  & \,\,\,\,\,\,2,008 (126) & \,\,\,\,\,\,1,254 (121) & \,\,\,\,\,\,1,283 (103) & \,\,\,\,\,\,\,\,--120 \,\,\,(75)\tabularnewline
--5 &  & \,\,\,\,\,\,1,550 (112) & \,\,\,\,\,\,1,073 (107) & \,\,\,\,\,\,1,404 \,\,\,(71) & \,\,\,\,\,\,\,\,--177 \,\,\,(72)\tabularnewline
--4 &  & \,\,\,\,\,\,1,152 \,\,\,(90) & \,\,\,\,\,\,\,\,\,\,802 \,\,\,(87) & \,\,\,\,\,\,\,\,\,\,979 \,\,\,(86) & \,\,\,\,\,\,\,\,--175 \,\,\,(57)\tabularnewline
--3 &  & \,\,\,\,\,\,1,211 \,\,\,(70) & \,\,\,\,\,\,\,\,\,\,839 \,\,\,(71) & \,\,\,\,\,\,1,404 \,\,\,(71) & \,\,\,\,\,\,\,\,\,\,\,287 \,\,\,(43)\tabularnewline
--2 &  & \,\,\,\,\,\,\,\,\,\,686 \,\,\,(60) & \,\,\,\,\,\,\,\,\,\,487 \,\,\,(61) & \,\,\,\,\,\,\,\,\,\,626 \,\,\,(61) & \,\,\,\,\,\,\,\,\,\,\,163 \,\,\,(42)\tabularnewline
--1 &  & {[}normalized to 0{]} & {[}normalized to 0{]} & {[}normalized to 0{]} & \,\,\,\,\,\,\,\,--120 \,\,\,(50)\tabularnewline
0 &  & \,\,\,--7,660 (164) & \,\,\,--7,339 (167) & \,\,\,--7,568 (165) & \,\,\,--5,839 (188)\tabularnewline
1 &  & --12,440 (249) & --11,647 (256) & --12,233 (251) & \,\,\,--9,799 (284)\tabularnewline
2 &  & --10,505 (241) & \,\,\,--9,319 (250) & --10,169 (244) & \,\,\,--7,044 (285)\tabularnewline
3 &  & \,\,\,--9,852 (240) & \,\,\,--8,405 (251) & \,\,\,--9,447 (244) & \,\,\,--5,571 (295)\tabularnewline
4 &  & \,\,\,--9,294 (246) & \,\,\,--7,744 (255) & \,\,\,--8,896 (249) & \,\,\,--4,192 (313)\tabularnewline
5 &  & \,\,\,--8,823 (251) & \,\,\,--7,233 (260) & \,\,\,--8,476 (254) & \,\,\,--2,900 (331)\tabularnewline
6 &  & \,\,\,--8,447 (255) & \,\,\,--6,856 (263) & \,\,\,--8,137 (258) & \,\,\,--1,701 (347)\tabularnewline
7 &  & \,\,\,--8,095 (258) & \,\,\,--6,504 (265) & \,\,\,--7,810 (261) & \,\,\,\,\,\,\,\,--526 (364)\tabularnewline
8 &  & \,\,\,--7,707 (260) & \,\,\,--6,123 (266) & \,\,\,--7,433 (263) & \,\,\,\,\,\,\,\,\,\,\,684 (482)\tabularnewline
9 &  & \,\,\,--7,554 (264) & \,\,\,--5,945 (270) & \,\,\,--7,265 (267) & \,\,\,\,\,\,1,659 (401)\tabularnewline
\hline 
\end{tabular}
\par\end{centering}
\begin{tablenotes}
\footnotesize

\item Note: Estimates are from doubly robust matching DID (columns
A--C; \citealp{sant2020doubly}) or a regression with unit-specific
trends (column D; equation (\ref{eq:unit-specific-trends})). All
specifications include age and calendar year indicators as covariates.
Column B additionally includes region, occupation, and tenure controls,
and column C instead adds lagged earnings. For column D, the normalization
imposes a zero-average condition ($\sum^{-1}_{k=-9}\beta_{k}=0$)
and a flat-trend condition ($\sum^{-1}_{k=-9}\beta_{k}(k+5)=0$) on
the pre-displacement coefficients. Standard errors (in parentheses)
are clustered at the individual worker level. Earnings are expressed
in year-2000 Euros.

\end{tablenotes}
\end{threeparttable}
\end{table}

\subsection{Normalization of Unobserved Heterogeneity\label{subsec:Normalization-of-Unobserved}}

The assumption of independent uniform unobserved heterogeneity $U_{i}=(U_{1i},U_{2i})$
serves as a normalization rather than a substantive distributional
restriction. I demonstrate that for any continuous bivariate distribution
of unobserved heterogeneity, the model can be equivalently represented
using independent uniform variables through appropriate transformations.

For simplicity, I suppress covariates $X_{i}$ in the exposition.
Consider an arbitrary continuous bivariate distribution of unobserved
heterogeneity $V_{i}=(V_{1i},V_{2i})$ with cumulative distribution
functions $F_{V_{1}}(v_{1})$ and $F_{V_{2}|V_{1}}(v_{2}|v_{1})$.
Following Assumption 4--(b), choose the normalization such that $M\left[f_{\boldsymbol{Y}^{\text{pre}}(0)|V}(\cdot|v)\right]=(v_{1},v_{2})$,
meaning that a certain measure of location (e.g., mean, mode, or quantile)
of pre-treatment outcomes determines the unobserved heterogeneity
$V_{i}$.

Consider the transformations $U_{1i}=F_{V_{1}}(V_{1i})$ and $U_{2i}=F_{V_{2}|V_{1}}(V_{2i}|V_{1i})$.
By construction, $U_{1i}$ follows a uniform distribution on $(0,1)$,
and $U_{2i}|U_{1i}$ also follows a uniform distribution on $(0,1)$.
The original unobserved heterogeneity can be recovered through the
inverse transformations
\begin{eqnarray}
V_{1i} & = & F^{-1}_{V_{1}}(U_{1i}),\\
V_{2i} & = & F^{-1}_{V_{2}|V_{1}}(U_{2i}|F^{-1}_{V_{1}}(U_{1i})).
\end{eqnarray}
Then, any function $\tilde{g}(V_{1i},V_{2i})$ in the original model
can be expressed as a function $g(U_{1i},U_{2i})$ in the transformed
model:
\begin{equation}
\tilde{g}(V_{1i},V_{2i})=g(U_{1i},U_{2i})\equiv\tilde{g}\left(F^{-1}_{V_{1}}(U_{1i}),F^{-1}_{V_{2}|V_{1}}(U_{2i}|F^{-1}_{V_{1}}(U_{1i}))\right).
\end{equation}
This applies to all conditional densities and the displacement probability
$\Pr\left[D_{i}=1|V_{i}\right]$ in the original model.

Therefore, the model with arbitrary bivariate $V_{i}$ and the model
with independent uniform $U_{i}$ are observationally equivalent;
they generate identical distributions for all observed variables.
Once the model with $U_{i}$ is identified, then the model with $V_{i}$
can also be identified. Thus, the specification with independent uniform
unobserved heterogeneity imposes no loss of generality and serves
purely as a convenient normalization.

In my empirical specification, I normalize unobserved heterogeneity
in the original model with $V_{i}$ by the location shifter of pre-treatment
outcomes: $\left(\tilde{\mu}_{-1}(v),\tilde{\mu}_{-2}(v)\right)=(v_{1},v_{2})$.
Thus, the location shifters in the transformed model with $U_{i}$
are
\begin{align}
\mu_{-1}(u_{1},u_{2}) & =F^{-1}_{V_{1}}(u_{1}),\\
\mu_{-2}(u_{1},u_{2}) & =F^{-1}_{V_{2}|V_{1}}(u_{2}|F^{-1}_{V_{1}}(u_{1})),
\end{align}
which are approximated using Chebyshev polynomials. Note that $\mu_{-1}$
depends only on $u_{1}$, while $\mu_{-2}$ depends on both $u_{1}$
and $u_{2}$. This structure reflects the original normalization:
since $\tilde{\mu}_{-1}(v)=v_{1}$ in the original model, it can depend
only on the first dimension even after transformation.

\subsection{Likelihood Function}

With the conditioning on covariates omitted for conciseness, the individual
likelihood is given by
\begin{multline}
\int^{1}_{0}\int^{1}_{0}\exp\biggl\{\ln f_{\ln Y_{-1}(0),\ln Y_{-2}(0)|U}\left(\ln Y_{i,-1},\ln Y_{i,-2}|u\right)+\ln f_{\ln Y_{0}(0)|U,D}\left(\ln Y_{i0}|u,D_{i}\right)\\
+D_{i}\ln\Pr[D_{i}=1|U_{i}=u]+(1-D_{i})\ln\Pr[D_{i}=0|U_{i}=u]\\
+(1-D_{i})(1-Z_{i1})(1-Z_{i2})\left(\ln\Pr[Z_{i1}=0,Z_{i2}=0|U_{i}=u]+\ln f_{\tilde{Y}_{1},\tilde{Y}_{2}|U,Z_{1},Z_{2}}\left(\ln Y_{i1},\ln Y_{i2}|u,0,0\right)\right)\\
+(1-D_{i})Z_{i1}(1-Z_{i2})\left(\ln\Pr[Z_{i1}=1,Z_{i2}=0|U_{i}=u]+\ln f_{\tilde{Y}_{2}|U,Z_{1},Z_{2}}\left(\ln Y_{i2}|u,1,0\right)\right)\\
+(1-D_{i})(1-Z_{i1})Z_{i2}\left(\ln\Pr[Z_{i1}=0,Z_{i2}=1|U_{i}=u]+\ln f_{\tilde{Y}_{1}|U,Z_{1},Z_{2}}\left(\ln Y_{i1}|u,0,1\right)\right)\\
+(1-D_{i})Z_{i1}Z_{i2}\ln\Pr[Z_{i1}=1,Z_{i2}=1|U_{i}=u]\biggr\} du_{1}du_{2}.
\end{multline}
The integral is numerically approximated using Gauss--Legendre quadrature
with 11 nodes in each dimension, which can integrate polynomials of
degree up to 21 exactly.

The following discussion focuses on the log-likelihood contribution
for post-treatment period earnings in detail; the contribution for
pre-treatment and reference period earnings follows analogously. The
functional form of the likelihood contribution depends on whether
earnings are top-coded or zero in each period. Let $C_{it}$ denote
the top-coding threshold for period $t$. The case of zero earnings
can be treated as equivalent to $\ln C_{it}=-\infty$. I define $\omega^{\text{post}}_{00}=1$
to simplify the exposition.

\paragraph{Earnings not top-coded and not zero in both periods}

When $Y_{i1}$ and $Y_{i2}$ are both positive and below their respective
thresholds, the log-likelihood contribution is
\begin{multline}
2\ln\left|\sum_{0\le j+k\le4}\frac{\omega^{\text{post}}_{jk}}{\sqrt{j!k!}}H_{j}\left(\frac{\ln Y_{i1}-\mu_{1}(u,0)}{\sigma_{1}(u,0)}\right)H_{k}\left(\frac{\ln Y_{i2}-\mu_{2}(u,0)}{\sigma_{2}(u,0)}\right)\right|-\frac{1}{2}\left(\frac{\ln Y_{i1}-\mu_{1}(u,0)}{\sigma_{1}(u,0)}\right)^{2}\\
-\frac{1}{2}\left(\frac{\ln Y_{i2}-\mu_{2}(u,0)}{\sigma_{2}(u,0)}\right)^{2}-\ln\sigma_{1}(u,0)-\ln\sigma_{2}(u,0)-\ln\sum_{0\le j+k\le4}\left(\omega^{\text{post}}_{jk}\right)^{2}-\ln(2\pi).
\end{multline}

\paragraph{Earnings top-coded or zero in one period, not in the other}

When earnings are top-coded or zero in one period but observed in
the other, the log-likelihood contribution involves integrating over
the unobserved dimension. For example, if $0<Y_{i1}<C_{i1}$ and $Y_{i2}=C_{i2}$,
the contribution is:
\begin{multline}
\ln\sum_{0\le j+k,\ell+m\le4}\frac{\omega^{\text{post}}_{jk}\omega^{\text{post}}_{\ell m}}{\sqrt{j!k!\ell!m!}}H_{j}\left(\frac{\ln Y_{i1}-\mu_{1}(u,0)}{\sigma_{1}(u,0)}\right)H_{\ell}\left(\frac{\ln Y_{i1}-\mu_{1}(u,0)}{\sigma_{1}(u,0)}\right)\int^{\infty}_{\frac{\ln C_{i2}-\mu_{2}(u,0)}{\sigma_{2}(u,0)}}\phi(v)H_{k}(v)H_{m}(v)dv\\
-\frac{1}{2}\left(\frac{\ln Y_{i1}-\mu_{1}(u,0)}{\sigma_{1}(u,0)}\right)^{2}-\ln\sigma_{1}(u,0)-\ln\sum_{0\le j+k\le4}\left(\omega^{\text{post}}_{jk}\right)^{2}-\frac{1}{2}\ln(2\pi),
\end{multline}
where $\omega^{\text{post}}_{00}=1$. The integral in this expression
has a closed form involving the standard normal density and cumulative
distribution function and can be computed efficiently.

\paragraph{Earnings top-coded or zero in both periods}

When both earnings are either top-coded or zero, the log-likelihood
contribution involves a double integral:
\begin{multline}
\ln\sum_{0\le j+k,\ell+m\le4}\frac{\omega^{\text{post}}_{jk}\omega^{\text{post}}_{\ell m}}{\sqrt{j!k!\ell!m!}}\int^{\infty}_{\frac{\ln C_{i1}-\mu_{1}(u,Z_{i2})}{\sigma_{1}(u,Z_{i2})}}\phi(v)H_{j}(v)H_{\ell}(v)dv\int^{\infty}_{\frac{\ln C_{i2}-\mu_{2}(u,Z_{i1})}{\sigma_{2}(u,Z_{i1})}}\phi(v)H_{k}(v)H_{m}(v)dv\\
-\ln\sum_{0\le j+k\le4}\left(\omega^{\text{post}}_{jk}\right)^{2}.
\end{multline}

\subsection{ATT Estimator and Its Standard Error\label{subsec:ATT-Estimator-and}}

Suppose that the estimated model yields the parameter vector $\hat{\psi}$
governing the joint densities of potential outcomes and treatment
status. Following equation (\ref{eq:ATT_x}), I estimate
\begin{equation}
\hat{\theta}^{\text{ATT}}_{t}=\hat{\theta}^{\text{M}}_{t}-\frac{\sum^{N}_{i=1}D_{i}b_{it}(\hat{\psi})}{\sum^{N}_{i=1}D_{i}},
\end{equation}
where 
\begin{equation}
b_{it}(\psi)=\int_{u\in{\cal U}}E_{\psi}\left[Y_{it}(0)|U_{i}=u,X_{i}\right]\left(f_{U|D=1,X}(u|X_{i};\psi)-f_{U|D=0,X}(u|X_{i};\psi)\right)du
\end{equation}
is the bias correction term implied by parameter $\psi$, which depends
on the model-implied conditional mean of the outcome and the conditional
distribution of unobservables. As in the likelihood function, the
model-implied conditional mean accounts for top-coding of earnings
at the social security contribution ceiling.

The matching estimator $\hat{\theta}^{\text{M}}_{t}$ admits an asymptotically
linear expansion:
\begin{equation}
\sqrt{N}\left(\hat{\theta}^{\text{M}}_{t}-\theta^{\text{M}}_{t}\right)=\frac{1}{\sqrt{N}}\sum^{N}_{i=1}a_{it}+o_{P}(1),
\end{equation}
where the exact form of the influence function $a_{it}$ is provided
by \citet{vermeulen2015bias} and \citet{sant2020doubly}.

\citet{ackerberg2012practical} demonstrate that a semiparametric
two-step estimator using a series approximation in the first step
can be treated as a parametric estimator for the purpose of standard
error computation. Standard asymptotics of parametric maximum likelihood
estimation imply:
\begin{equation}
\sqrt{N}\left(\hat{\psi}-\psi\right)=-E\left[\nabla s_{i}(\psi)\right]^{-1}\frac{1}{\sqrt{N}}\sum^{N}_{i=1}s_{i}(\psi)+o_{P}(1),
\end{equation}
where $s_{i}(\psi)$ is the score function.

The estimator $\hat{\theta}^{\text{ATT}}_{t}$ therefore admits an
asymptotically linear expansion:
\begin{multline}
\sqrt{N}\left(\hat{\theta}^{\text{ATT}}_{t}-\theta^{\text{ATT}}_{t}\right)=\frac{1}{\sqrt{N}}\sum^{N}_{i=1}\Bigl\{ a_{it}-\frac{D_{i}}{E[D_{i}]}\left(b_{it}(\psi)-\frac{E[D_{i}b_{it}(\psi)]}{E[D_{i}]}\right)\\
+\frac{E[D_{i}\nabla b_{it}(\psi)]'}{E[D_{i}]}E\left[\nabla s_{i}(\psi)\right]^{-1}s_{i}(\psi)\Bigr\}+o_{P}(1).
\end{multline}
I use this asymptotic expansion to compute standard errors clustered
at the individual worker level.

\subsection{Estimates with Different Year-Pair Choices\label{subsec:Point-Estimates-with-Year-Pair}}

Table \ref{Table:ATT_full_alt} presents the exact point estimates
and standard errors corresponding to the graphical results shown in
Figure \ref{Fig:Alt}. Each row reports estimates for a specific year
after displacement. Column (1) gives the standard DID estimates for
reference. Column (2) reports the ATT estimates from the alternative
approach, obtained using the pair of post-treatment observations listed
in the fourth column. Column (3) reports the estimate for the other
year in that pair, from the same specification.

\begin{table}
\caption{The Estimated Earnings Losses by Year After Displacement}

\label{Table:ATT_full_alt}

\centering
\begin{threeparttable}
\begin{centering}
\begin{tabular}{ccccc}
 &  &  &  & \tabularnewline
\hline 
\multirow{2}{*}{Year} & Standard DID & Alternative Method & Paired Year Est. & Obs. Pair\tabularnewline
 & (1) & (2) & (3) & (4)\tabularnewline
\hline 
0 & \,\,\,--7,660 (164) & \,\,\,--6,443 (276) & \,\,\,--4,482 (389) & (0, 9)\tabularnewline
1 & --12,440 (249) & --11,013 (367) & \,\,\,--5,150 (426) & (1, 9)\tabularnewline
2 & --10,505 (241) & \,\,\,--8,547 (401) & \,\,\,--5,048 (454) & (2, 9)\tabularnewline
3 & \,\,\,--9,852 (240) & \,\,\,--7,158 (447) & \,\,\,--4,465 (486) & (3, 9)\tabularnewline
4 & \,\,\,--9,294 (246) & \,\,\,--6,370 (463) & \,\,\,--4,312 (496) & (4, 9)\tabularnewline
5 & \,\,\,--8,823 (251) & \,\,\,--5,851 (525) & \,\,\,--4,221 (570) & (5, 9)\tabularnewline
6 & \,\,\,--8,447 (255) & \,\,\,--5,528 (551) & \,\,\,--4,307 (597) & (6, 9)\tabularnewline
7 & \,\,\,--8,095 (258) & \,\,\,--4,741 (589) & \,\,\,--3,891 (646) & (7, 9)\tabularnewline
8 & \,\,\,--7,707 (260) & \,\,\,--4,234 (586) & \,\,\,--5,645 (536) & (8, 5)\tabularnewline
9 & \,\,\,--7,554 (264) & \,\,\,--4,221 (570) & \,\,\,--5,851 (525) & (9, 5)\tabularnewline
\hline 
\end{tabular}
\par\end{centering}
\begin{tablenotes}
\footnotesize

\item Note: Column (1) reports the standard DID estimates from Panel
A of Figure \ref{Fig:DID}. Column (2) reports estimates obtained
by re-estimating the model using the pairs of post-treatment observations
listed in column (4). Column (3) reports the estimate for the other
year in each pair, from the same estimation. Standard errors (in parentheses)
are clustered at the individual worker level. Earnings are expressed
in year-2000 Euros.

\end{tablenotes}
\end{threeparttable}
\end{table}

The year-9 effect is estimated in every specification that pairs year
9 with years 0 through 7; these estimates range from --3,891 to --5,150
Euros, roughly 52 to 68\% of the standard DID estimate. While being
broadly robust to the choice of intermediate post-treatment period,
the estimates tend to move toward the DID benchmark when year 9 is
paired with earlier intermediate years. This pattern raises the concern
of serial correlation between those years' outcomes and the reference
period outcome beyond what $U_{i}$ captures (e.g., through moving-average
shocks), which would imply a violation of Assumption 1. At the same
time, standard errors are larger with later intermediate years, since
observations close to year 9 carry less independent information about
distinct dimensions of $U_{i}$. The baseline choice of years 5 and
9 balances these considerations, guarding against a potential violation
of Assumption 1 at a modest cost in precision.

\section{Three-Dimensional Specification\label{Asec:Three-Dimensional-Specification}}

This section specifies and estimates a version of the model in Section
\ref{subsec:Estimation-Without-Parallel} with three-dimensional unobserved
heterogeneity, $U_{i}=(U_{1i},U_{2i},U_{3i})$, where each dimension
is independently drawn from a uniform distribution on $(0,1)$. As
in the two-dimensional case, this distributional specification serves
as a normalization rather than a substantive assumption, by the argument
given in Appendix \ref{subsec:Normalization-of-Unobserved}. The sample
and the estimation strategy are the same as in Section \ref{subsec:Estimation-Without-Parallel}.

The identification results in Section \ref{sec:Econometric-Framework}
require $K\le\min\{T_{\text{pre}},T_{\text{post}}\}$, so I set $T_{\text{pre}}=T_{\text{post}}=3$,
using three pre-treatment and three post-treatment observations. I
use earnings from three, six, and nine years before displacement as
the three pre-treatment outcomes, and earnings from one, five, and
nine years after displacement as the three post-treatment outcomes.
This choice balances two considerations. The first is separation from
the reference period. The SIAB data are spell-based (Appendix \ref{subsec:Sample-Construction-Details}),
and a spell spanning two consecutive calendar years contributes to
the constructed earnings of both, so adjacent years' earnings can
be mechanically correlated. More generally, moving-average shocks
would induce serial correlation between outcomes beyond what $U_{i}$
captures, in violation of Assumption 1. The nearest pre- and post-treatment
outcomes are therefore placed two years away from the reference period.
The second is separation within each block, so that the observations
carry independent information about each dimension of $U_{i}$. The
post-treatment block has the same four-year separation as the two-dimensional
baseline; the pre-treatment block is tighter, at three years, to accommodate
the two-year gap from the reference period.

\subsection{Empirical Specification\label{Asubsec:Three-Dimensional-Empirical-Specification}}

As in Section \ref{subsec:Estimation-Without-Parallel}, every scalar
function of $(u,x)$ in the model is partially linear, entering nonlinearly
in $u$ through Chebyshev polynomials and linearly in $x$. With three
dimensions of unobserved heterogeneity the common form is 
\begin{equation}
\sum_{0\le j+k+l\le2}\alpha_{jkl}T_{j}(u_{1})T_{k}(u_{2})T_{l}(u_{3})+x'\beta,\label{eq:partially_linear_3d}
\end{equation}
where the total degree of the Chebyshev basis is lowered from three
to two so that adding a third dimension does not enlarge the basis.

The probability of displacement is specified as in equation (\ref{eq:prob_treated}),
with the third Chebyshev index added.

The conditional density of the three pre-treatment outcomes is specified
as: 
\begin{equation}
f_{\ln Y_{-1}(0),\ln Y_{-2}(0),\ln Y_{-3}(0)|U,X}\left(y_{-1},y_{-2},y_{-3}|u,x\right)=\frac{h_{3}\left(\frac{y_{-1}-\mu_{-1}(u,x)}{\sigma_{-1}(u,x)},\frac{y_{-2}-\mu_{-2}(u,x)}{\sigma_{-2}(u,x)},\frac{y_{-3}-\mu_{-3}(u,x)}{\sigma_{-3}(u,x)};\boldsymbol{\omega}^{\text{pre}}\right)}{\sigma_{-1}(u,x)\sigma_{-2}(u,x)\sigma_{-3}(u,x)},\label{eq:pre_density_3d}
\end{equation}
where $y_{-1}$, $y_{-2}$, and $y_{-3}$ represent log earnings from
three, six, and nine years before displacement ($t=-1$, $t=-2$,
and $t=-3$), respectively. The function $h_{3}$ is the three-dimensional
counterpart of the sieve approximation in equation (\ref{eq:density_hermite}):
\begin{equation}
h_{3}\left(v_{1},v_{2},v_{3};\boldsymbol{\omega}\right)=\frac{\phi(v_{1})\phi(v_{2})\phi(v_{3})\left(1+\sum_{1\le j+k+l\le3}\omega_{jkl}\frac{H_{j}(v_{1})}{\sqrt{j!}}\frac{H_{k}(v_{2})}{\sqrt{k!}}\frac{H_{l}(v_{3})}{\sqrt{l!}}\right)^{2}}{1+\sum_{1\le j+k+l\le3}\omega^{2}_{jkl}}.\label{eq:density_hermite_3d}
\end{equation}
As with the Chebyshev basis, the sieve is truncated at a lower total
degree than in the two-dimensional specification, three rather than
four, to limit the increase in the number of parameters. The location
functions $\mu_{t}(u,x)$ and the log scale functions $\ln\sigma_{t}(u,x)$
take the form of equation (\ref{eq:partially_linear_3d}), while the
higher-order features captured by $\boldsymbol{\omega}^{\text{pre}}$
are held constant across $(u,x)$.

The reference period requires a different treatment. Assumption 4--(c)
demands that distinct values of $U_{i}$ induce distinct conditional
densities of $Y_{i0}(0)$, so with $K=3$ at least three features
of the density must vary with $U_{i}$. The two-dimensional specification
holds the sieve shape parameter $\boldsymbol{\omega}^{0}$ constant
across $u$, leaving only the location $\mu_{0}(u,x,d)$ and the scale
$\sigma_{0}(u,x,d)$, one function short. Letting the shape parameter
vary with $u$ would close the gap, but doing so for the whole vector
$\boldsymbol{\omega}^{0}$ adds a large number of parameters, and
picking one coefficient to depend on $u$ is arbitrary. I therefore
use the sinh-arcsinh family of \citet{jones2009sinh}, letting its
skewness parameter depend on $u$.

For $\nu\in\mathbb{R}$ and $\delta>0$, the family is generated by
applying the sinh-arcsinh transformation to a standard normal variable
$\xi$: 
\begin{equation}
\ln Y_{i0}(0)=\mu_{0}(u,x,d)+\sigma_{0}(u,x,d)\sinh\left(\frac{\operatorname{asinh}(\xi)+\nu(u,x,d)}{\delta}\right).\label{eq:shash_representation}
\end{equation}
The parameter $\nu$ controls asymmetry, with $\nu>0$ skewing the
distribution to the right, and $\delta$ controls tail weight, with
$\delta<1$ giving heavier tails than the normal and $\delta>1$ lighter
tails. The standard normal distribution is nested at $\nu=0$ and
$\delta=1$.

I specify the functions $\mu_{0}(u,x,d)$, $\ln\sigma_{0}(u,x,d)$,
and $\nu(u,x,d)$ to take the form of equation (\ref{eq:partially_linear_3d}).
As in Section \ref{subsec:Estimation-Without-Parallel}, these functions
depend on treatment status $d$ in addition to $(u,x)$, consistent
with Assumption 1, through additive shifts.

As in the two-dimensional specification, post-treatment untreated
outcomes are modeled as $Y_{it}(0)=(1-Z_{it})\exp(\tilde{Y}_{it})$
for $t=1,2,3$, where $Z_{it}$ is an indicator for zero earnings
in period $t$ and $\tilde{Y}_{it}$ denotes log earnings. The conditional
probability $\Pr[Z_{i1}=1|U_{i},X_{i}]$ follows the partially linear
specification in equation (\ref{eq:partially_linear_3d}). The conditional
probabilities $\Pr[Z_{i2}=1|U_{i},X_{i},Z_{i1}]$ and $\Pr[Z_{i3}=1|U_{i},X_{i},Z_{i1},Z_{i2}]$
are specified analogously, with $U_{i}$ entering nonlinearly via
Chebyshev polynomials and $(X_{i},Z_{i1})$ and $(X_{i},Z_{i1},Z_{i2})$
entering linearly. The conditional density of log earnings is given
by: 
\begin{equation}
f_{\tilde{Y}_{1},\tilde{Y}_{2},\tilde{Y}_{3}|U,X,Z_{1},Z_{2},Z_{3}}\left(y_{1},y_{2},y_{3}|u,x,z_{1},z_{2},z_{3}\right)=\frac{h_{3}\left(\frac{y_{1}-\mu_{1}(u,x,z_{2},z_{3})}{\sigma_{1}(u,x,z_{2},z_{3})},\frac{y_{2}-\mu_{2}(u,x,z_{1},z_{3})}{\sigma_{2}(u,x,z_{1},z_{3})},\frac{y_{3}-\mu_{3}(u,x,z_{1},z_{2})}{\sigma_{3}(u,x,z_{1},z_{2})};\boldsymbol{\omega}^{\text{post}}\right)}{\sigma_{1}(u,x,z_{2},z_{3})\sigma_{2}(u,x,z_{1},z_{3})\sigma_{3}(u,x,z_{1},z_{2})},\label{eq:post_density_3d}
\end{equation}
where $h_{3}$ is defined in equation (\ref{eq:density_hermite_3d}).
The location and scale functions are nonlinear in $u$ and linear
in $(x,z)$. Since $\tilde{Y}_{it}$ affects the actual outcome $Y_{it}(0)$
only when $Z_{it}=0$, the specification allows the location and scale
to be shifted by the zero-earnings indicators in the other two periods
only.

The resulting specification has 274 parameters, a modest increase
over the 189 parameters of the two-dimensional specification. The
integral over ${\cal U}$ is approximated by product Gauss--Legendre
quadrature with seven nodes in each dimension, rather than eleven
nodes in each of two dimensions, so the number of grid points increases
from $11^{2}=121$ to $7^{3}=343$, keeping the computational costs
manageable.

\subsection{Estimation Results\label{Asubsec:Three-Dimensional-Estimation-Results}}

Table \ref{Table:ATT_K3} presents the estimated effects of job displacement
on earnings one, five, and nine years after displacement. Column (1)
gives the standard DID estimates for reference. Column (2) reports
the ATT estimates from the three-dimensional specification, and column
(3) the difference between the two estimates.

\begin{table}
\caption{The Estimated Earnings Losses from Displacement ($K=3$)}

\label{Table:ATT_K3}

\centering
\begin{threeparttable}
\begin{centering}
\begin{tabular}{ccccc}
 &  &  &  & \tabularnewline
\hline 
Years &  & Standard & Alternative: & Difference:\tabularnewline
from &  & DID: $\hat{\theta}^{\text{DID,M}}_{t}$ & $\hat{\theta}^{\text{ATT}}_{t}$ & $\hat{\theta}^{\text{DID,M}}_{t}-\hat{\theta}^{\text{ATT}}_{t}$\tabularnewline
Displacement &  & (1) & (2) & (3)\tabularnewline
\cline{1-1}\cline{3-5}
\multirow{2}{*}{1} &  & --12,440 & --10,234 & --2,206\tabularnewline
 &  & \,\,\,\,\,\,\,\,(249) & \,\,\,\,\,\,\,\,(336) & \,\,\,\,\,\,(231)\tabularnewline
\multirow{2}{*}{5} &  & \,\,--8,823 & \,\,--4,366 & --4,457\tabularnewline
 &  & \,\,\,\,\,\,\,\,(251) & \,\,\,\,\,\,\,\,(409) & \,\,\,\,\,\,(332)\tabularnewline
\multirow{2}{*}{9} &  & \,\,--7,554 & \,\,--4,326 & --3,228\tabularnewline
 &  & \,\,\,\,\,\,\,\,(264) & \,\,\,\,\,\,\,\,(389) & \,\,\,\,\,\,(303)\tabularnewline
\hline 
\end{tabular}
\par\end{centering}
\begin{tablenotes}
\footnotesize

\item Note: Standard DID estimates correspond to Panel A of Figure
\ref{Fig:DID}. Standard errors (in parentheses) are clustered at
the individual worker level. Outcomes are measured by annual earnings
in year-2000 Euros.

\end{tablenotes}
\end{threeparttable}
\end{table}

The three-dimensional estimates are substantially smaller in magnitude
than the standard DID estimates at the medium and long horizons: 4,366
Euros five years after displacement, 49\% of the standard DID estimate
of 8,823 Euros, and 4,326 Euros nine years after displacement, 57\%
of 7,554 Euros. One year after displacement the gap is narrower, with
the estimate of 10,234 Euros amounting to 82\% of the DID estimate.

\section{Estimation of Treatment Effect Heterogeneity}

This section implements the extension suggested in Section \ref{subsec:Identifying-Treatment-Effect}
to identify treatment effect heterogeneity. The framework allows recovery
of the $(U_{i},X_{i})$-conditional average treatment effect $E[Y_{it}(1)-Y_{it}(0)|U_{i},X_{i}]$,
which in turn enables identification of the average treatment effect
on the untreated (ATU), the population average treatment effect (ATE),
and the $U_{i}$-driven component of treatment effect variance.

The estimation extends the baseline approach by incorporating the
conditional distribution of treated outcomes $\boldsymbol{Y}^{\text{post}}_{i}(1)$
into the likelihood function: 
\begin{multline}
\sum^{N}_{i=1}\ln\int_{u\in{\cal U}}f_{\boldsymbol{Y}^{\text{pre}}(0)|U,X}\bigl(\boldsymbol{Y}^{\text{pre}}_{i}|u,X_{i}\bigr)\,f_{Y_{0}(0),D|U,X}\bigl(Y_{i0},D_{i}|u,X_{i}\bigr)\\
\times f_{\boldsymbol{Y}^{\text{post}}(0)|U,X}\bigl(\boldsymbol{Y}^{\text{post}}_{i}|u,X_{i}\bigr)^{1-D_{i}}\,f_{\boldsymbol{Y}^{\text{post}}(1)|U,X}\bigl(\boldsymbol{Y}^{\text{post}}_{i}|u,X_{i}\bigr)^{D_{i}}\,f_{U}(u)\,du.
\end{multline}
The treated-outcome density $f_{\boldsymbol{Y}^{\text{post}}(1)|U,X}$
is specified to parallel its untreated counterpart $f_{\boldsymbol{Y}^{\text{post}}(0)|U,X}$
in Section \ref{subsec:Estimation-Without-Parallel} in every respect,
including the zero-earnings mixture, the squared-Hermite sieve density
of log earnings, and the partially linear forms of the location and
scale functions.

Using the estimated conditional densities, I construct the model-implied
average treatment effect conditional on $(U,X)$: 
\begin{equation}
\hat{\delta}_{t}(u,x)\equiv\hat{E}[Y_{it}(1)|U_{i}=u,X_{i}=x]-\hat{E}[Y_{it}(0)|U_{i}=u,X_{i}=x],
\end{equation}
and combine it with empirical sums to form 
\begin{align}
\hat{\theta}^{\text{ATT}}_{t}\, & =\,\frac{1}{\sum^{N}_{i=1}D_{i}}\sum^{N}_{i=1}D_{i}\!\int\hat{\delta}_{t}(u,X_{i})\,\hat{f}_{U|D=1,X}(u|X_{i})\,du,\\
\hat{\theta}^{\text{ATU}}_{t}\, & =\,\frac{1}{\sum^{N}_{i=1}(1-D_{i})}\sum^{N}_{i=1}(1-D_{i})\!\int\hat{\delta}_{t}(u,X_{i})\,\hat{f}_{U|D=0,X}(u|X_{i})\,du,\\
\hat{\theta}^{\text{ATE}}_{t}\, & =\,\frac{1}{N}\sum^{N}_{i=1}\hat{m}^{\text{CATE}}_{t}(X_{i}),\\
\hat{\theta}^{\text{SD-U}}_{t}\, & =\,\sqrt{\frac{1}{N}\sum^{N}_{i=1}\!\int\bigl(\hat{\delta}_{t}(u,X_{i})-\hat{m}^{\text{CATE}}_{t}(X_{i})\bigr)^{2}du},\\
\hat{\theta}^{\text{SD-X}}_{t}\, & =\,\sqrt{\frac{1}{N}\sum^{N}_{i=1}\bigl(\hat{m}^{\text{CATE}}_{t}(X_{i})-\hat{\theta}^{\text{ATE}}_{t}\bigr)^{2}},
\end{align}
where $\hat{m}^{\text{CATE}}_{t}(X_{i})\equiv\int\hat{\delta}_{t}(u,X_{i})\,du$
is the within-$X$ conditional average treatment effect (CATE) evaluated
at $X_{i}$. Computation of standard errors follows the asymptotic
expansion approach described in Appendix \ref{subsec:ATT-Estimator-and}
for the baseline ATT estimator.

The first three estimators target the standard treatment effect parameters:
ATT, ATU, and ATE. The last two, $\hat{\theta}^{\text{SD-U}}_{t}$
and $\hat{\theta}^{\text{SD-X}}_{t}$, target heterogeneity in treatment
effects. In particular, from the law of total variance the treatment
effect variance decomposes as: 
\begin{equation}
\mathrm{Var}\bigl(Y_{it}(1)-Y_{it}(0)\bigr)=E\bigl[\mathrm{Var}(Y_{it}(1)-Y_{it}(0)|U,X)\bigr]+E\bigl[\mathrm{Var}(\delta_{t}(U,X)|X)\bigr]+\mathrm{Var}\bigl({m}^{\text{CATE}}_{t}(X)\bigr).
\end{equation}
The first term cannot be identified without assumptions on the joint
distribution of $\bigl(Y_{it}(1),Y_{it}(0)\bigr)$, which this framework
leaves unrestricted. The second term, whose square root is targeted
by $\hat{\theta}^{\text{SD-U}}_{t}$, is the contribution of unobserved
heterogeneity to treatment effect variation, and is the novel object
this framework makes identifiable. The third term, whose square root
is targeted by $\hat{\theta}^{\text{SD-X}}_{t}$, reflects observable-driven
heterogeneity. This term is also recoverable through standard CATE
methods; it is estimated here for completeness.

Table~\ref{Table:TE} presents treatment effect estimates at five
and nine years after displacement. The ATU exceeds the ATT in absolute
magnitude at both horizons. This pattern likely reflects that untreated
workers have higher earnings and thus more to lose from displacement.
The ATE closely tracks the ATU due to the small treatment share. At
both horizons, $\hat{\theta}^{\text{SD-U}}_{t}$ is of similar magnitude
to the average treatment effect parameters, implying that unobserved
heterogeneity generates considerable dispersion in treatment effects.
By contrast, $\hat{\theta}^{\text{SD-X}}_{t}$, which captures the
heterogeneity attributable to observed characteristics, is substantially
smaller at both horizons.

\begin{table}
\caption{Extended Treatment Effect Estimates}

\label{Table:TE}

\centering
\begin{threeparttable} 

\begin{tabular}{cccccc}
 &  &  &  &  & \tabularnewline
\hline 
Years from & \multirow{2}{*}{ATT} & \multirow{2}{*}{ATU} & \multirow{2}{*}{ATE} & \multirow{2}{*}{$\hat{\theta}^{\text{SD-U}}_{t}$} & \multirow{2}{*}{$\hat{\theta}^{\text{SD-X}}_{t}$}\tabularnewline
displacement &  &  &  &  & \tabularnewline
\hline 
\multirow{2}{*}{5} & \,\,\,--6,485 & --10,315 & --10,214 & 7,166 & 567\tabularnewline
 & \,\,\,\,\,\,\,\,\,(407) & \,\,\,\,\,(1,042) & \,\,\,\,\,(1,018) & \,\,\,(872) & (271)\tabularnewline
\multirow{2}{*}{9} & \,\,\,--4,638 & \,\,\,--8,978 & \,\,\,--8,864 & 7,146 & 627\tabularnewline
 & \,\,\,\,\,\,(441) & \,\,\,\,\,(1,101) & \,\,(1,077) & (1,127) & (280)\tabularnewline
\hline 
\end{tabular}

\begin{tablenotes}
\footnotesize

\item Note: Standard errors (in parentheses) are clustered at the
individual worker level. Earnings are expressed in year-2000 Euros.

\end{tablenotes}
\end{threeparttable}
\end{table}

\section{Monte Carlo Simulations}

I investigate the finite-sample properties of the proposed estimator
through Monte Carlo simulations. Ideally, the data-generating process
(DGP) would mirror the specification employed in the empirical application,
so that simulated samples resemble the actual data. The full model,
however, is computationally too costly to re-estimate as many times
as a Monte Carlo study requires. Therefore, I simulate from a simplified
version that preserves the model's essential structure at lower estimation
costs. 

The simplified model omits top-coding and zero earnings, reduces the
covariates to a constant, and lowers the order of the Chebyshev and
Hermite polynomials by one. I take log earnings as the outcome: without
top-coding the earnings distribution is unbounded above, so a level
outcome would be sensitive to the upper tail, and the absence of zeros
keeps the log well-defined. I fit this simplified model to the same
data used in the empirical application, with the two post-treatment
periods $t=1$ and $t=2$ corresponding to five and nine years after
displacement. I treat the resulting estimates as the true DGP. Because
the model specifies only the untreated-outcome process, I complete
the design by setting $Y_{it}(1)=Y_{it}(0)$, so the true ATT is zero
by construction. I estimate the same simplified model on each simulated
sample, so the design is correctly specified.

Table \ref{Table:MC} reports results from 2,000 simulations at each
of five sample sizes. The proposed estimator shows minimal bias, while
the standard DID estimator is sizably biased. Neither bias depends
much on the sample size. The proposed estimator is noisier despite
its smaller bias; it trails the standard DID estimator in root mean
squared error at the smaller sample sizes but pulls ahead as the sample
size grows. Where this crossover falls is specific to the present
design: a larger or smaller DID bias in the DGP would move the break-even
point accordingly. Coverage for the proposed estimator approaches
the nominal 95\% level as the sample size grows, while DID coverage
deteriorates.

\begin{table}
\caption{Simulation Results}

\label{Table:MC}

\centering
\begin{threeparttable}

\begin{tabular}{ccccccccccc}
 &  & \multicolumn{4}{c}{} &  & \multicolumn{4}{c}{}\tabularnewline
\multicolumn{11}{l}{\textit{Panel A. Five years after displacement} ($t=1$)}\tabularnewline
\hline 
\multirow{2}{*}{\begin{cellvarwidth}[t]
\centering
Sample

Size
\end{cellvarwidth}} &  & \multicolumn{4}{c}{Proposed Estimator $\hat{\theta}^{\text{ATT}}_{t}$} &  & \multicolumn{4}{c}{Standard DID}\tabularnewline
\cline{3-6}\cline{8-11}
 &  & Bias & SD & rMSE & Cov. &  & Bias & SD & rMSE & Cov.\tabularnewline
\cline{1-1}\cline{3-6}\cline{8-11}
\,\,\,1,000 &  & --0.002 & 0.212 & 0.212 & 0.875 &  & --0.045 & 0.153 & 0.160 & 0.947\tabularnewline
\,\,\,2,000 &  & --0.008 & 0.153 & 0.153 & 0.896 &  & --0.048 & 0.111 & 0.121 & 0.943\tabularnewline
\,\,\,5,000 &  & --0.006 & 0.100 & 0.100 & 0.910 &  & --0.047 & 0.070 & 0.084 & 0.914\tabularnewline
10,000 &  & --0.003 & 0.068 & 0.068 & 0.927 &  & --0.047 & 0.050 & 0.069 & 0.859\tabularnewline
20,000 &  & --0.004 & 0.044 & 0.044 & 0.950 &  & --0.047 & 0.033 & 0.057 & 0.745\tabularnewline
\hline 
 &  & \multicolumn{4}{c}{} &  & \multicolumn{4}{c}{}\tabularnewline
\multicolumn{11}{l}{\textit{Panel B. Nine years after displacement} ($t=2$)}\tabularnewline
\hline 
\multirow{2}{*}{\begin{cellvarwidth}[t]
\centering
Sample

Size
\end{cellvarwidth}} &  & \multicolumn{4}{c}{Proposed Estimator $\hat{\theta}^{\text{ATT}}_{t}$} &  & \multicolumn{4}{c}{Standard DID}\tabularnewline
\cline{3-6}\cline{8-11}
 &  & Bias & SD & rMSE & Cov. &  & Bias & SD & rMSE & Cov.\tabularnewline
\cline{1-1}\cline{3-6}\cline{8-11}
\,\,\,1,000 &  & \,\,\,0.000 & 0.390 & 0.390 & 0.888 &  & --0.087 & 0.241 & 0.257 & 0.939\tabularnewline
\,\,\,2,000 &  & --0.006 & 0.262 & 0.262 & 0.899 &  & --0.081 & 0.168 & 0.187 & 0.948\tabularnewline
\,\,\,5,000 &  & --0.011 & 0.174 & 0.175 & 0.902 &  & --0.088 & 0.110 & 0.141 & 0.885\tabularnewline
10,000 &  & --0.008 & 0.122 & 0.123 & 0.924 &  & --0.088 & 0.076 & 0.116 & 0.799\tabularnewline
20,000 &  & --0.008 & 0.080 & 0.080 & 0.947 &  & --0.087 & 0.052 & 0.101 & 0.634\tabularnewline
\hline 
\end{tabular}

\begin{tablenotes}
\footnotesize

\item Note: Each row summarizes 2,000 Monte Carlo replications of
a calibrated data-generating process in which the true ATT is zero.
Reported are the bias, standard deviation (SD), root mean squared
error (rMSE), and empirical coverage (Cov.) of the nominal 95\% confidence
interval for the proposed estimator $\hat{\theta}^{\text{ATT}}_{t}$
and the standard DID estimator.

\end{tablenotes}
\end{threeparttable}
\end{table}

\section{Applicability of Miao et al. (2018) to a DID Design\label{sec:Adapting-the-Double-Proxy}}

The identification strategy of this paper builds on \citet{hu2008instrumental}
by using $\boldsymbol{Y}^{\text{pre}}_{i}(0)$, $Y_{i0}(0)$, and
$\boldsymbol{Y}^{\text{post}}_{i}(0)$ as three proxies for $U_{i}$
that are conditionally independent. By contrast, \citet{deaner2023proxy}
follows \citet{miao2018identifying}, which identifies a confounded
causal effect from a different set of independence restrictions. In
the panel results of \citet{deaner2023proxy}, treatment may change
from period to period, and the period-$t$ effect is identified by
conditioning on treatment status in the preceding period, leaving
variation only in the first post-treatment period in a DID design.
This section instead considers the natural mapping of the approach
of \citet{miao2018identifying} to the DID design of Section \ref{subsec:Setup-and-Assumptions},
where treatment is binary and changes only once. I show that the mapping
identifies the ATT when the unobserved heterogeneity is time-invariant
but not when it evolves over time.

In the framework of \citet{miao2018identifying}, the causal effect
of $X$ on $Y$ is confounded by an unobserved $U$ and is identified
from two proxies $W$ and $Z$ satisfying
\begin{equation}
W\perp(Z,X)\,|\,U,\qquad Z\perp Y\,|\,(U,X),\label{eq:double-proxy}
\end{equation}
together with completeness conditions under which each proxy is informative
enough about $U$ to invert the relevant integral equation. Specialize
(\ref{eq:double-proxy}) to the DID design of Section \ref{subsec:Setup-and-Assumptions}:
the confounder $U$ is the latent heterogeneity $U_{i}$, the treatment
$X$ is the group indicator $D_{i}$, the outcome $Y$ is $Y_{it}$
for $t>0$, and the two proxies are $W=\boldsymbol{Y}^{\text{pre}}_{i}$
and $Z=\boldsymbol{Y}^{\text{post}}_{i}\setminus\{Y_{it}\}$. 

As stated, the conditional independence restrictions of \citet{deaner2023proxy},
which adapts (\ref{eq:double-proxy}) to a potential outcome framework,
apply to both $Y_{it}(0)$ and $Y_{it}(1)$, and the second proxy
$Z$ depends on both $\boldsymbol{Y}^{\text{post}}_{i}(0)$ and $\boldsymbol{Y}^{\text{post}}_{i}(1)$;
these restrictions therefore constrain treated outcomes, which in
turn constrain treatment effect heterogeneity. Identification under
restrictions on untreated potential outcomes alone, which Section
\ref{subsec:Identification} establishes within my framework, is not
achieved there.

Identifying the counterfactual $E[Y_{it}(0)|D_{i}=1]$, however, does
not in fact require restrictions on treated outcomes. A closer reading
of the proof of \citet[Theorem 1]{deaner2023proxy} reveals that the
following weaker conditions suffice:\begin{enumerate}[label=(\alph*)]

\item $(W,Y_{it}(0))\perp D_{i}\,|\,U_{i}$;

\item $Z\perp W\,|\,U_{i},D_{i}=0$;

\item $Z\perp Y_{it}(0)\,|\,U_{i},D_{i}=0$;

\item common support of $U_{i}$ across $D_{i}=1$ and $D_{i}=0$;

\item completeness of $W$ for $U_{i}$ given $D_{i}=0$;

\item completeness of $Z$ for $U_{i}$ given $D_{i}=0$;

\end{enumerate}together with a regularity condition under which one
of the completeness conditions (e) and (f) yields the bridge functions.
None of the conditions (a)--(f) restricts treated outcomes: the only
potential outcome that appears is $Y_{it}(0)$, and the proxy $Z$
is relevant only for the control group $D_{i}=0$, where it coincides
with the untreated outcomes.

\paragraph{Time-invariant heterogeneity}

The independence conditions (a) and (b) follow from Assumption 1.
The common support condition (d) follows from Assumption 2. The completeness
conditions (e) and (f) are of the kind the present framework already
maintains (Assumptions 3--4). The one requirement stronger than Assumption
1 is the condition (c), which requires that the period-$t$ untreated
outcome be independent of the other post-treatment untreated outcomes
given $(U_{i},D_{i}=0)$. Assumption 1 constrains dependence only
across pre-treatment, reference, and post-treatment blocks and leaves
the dependence within post-treatment outcomes unrestricted. Nevertheless,
the nonparametric fixed effects model in equation (\ref{eq:nonpara-FE})
satisfies both the condition (c) and Assumption 1; with $Y_{it}(0)=g_{0t}(U_{i},\varepsilon_{it})$
and $\varepsilon_{it}$ independent across periods given $U_{i}$,
the untreated post-treatment outcomes are mutually independent conditional
on $U_{i}$. Thus, the \citet{miao2018identifying} framework identifies
the ATT under this model, as does the \citet{hu2008instrumental}
framework employed in this paper.

\paragraph{Time-varying heterogeneity}

Suppose instead that the unobserved state evolves, as in the hidden
Markov model in equation (\ref{eq:hidden_Markov}). With $U_{i}=\eta_{i0}$,
Assumption 1 holds within my framework, and both $\boldsymbol{Y}^{\text{pre}}_{i}(0)$
and $\boldsymbol{Y}^{\text{post}}_{i}(0)$ can serve as informative
proxies for $U_{i}=\eta_{i0}$ to meet the completeness conditions.
By contrast, the mapping of \citet{miao2018identifying} must isolate
$Y_{it}(0)$ from the remaining post-treatment outcomes, and no choice
of $U_{i}$ satisfies both the independence condition (c) and completeness:
\begin{itemize}
\item With $U_{i}=\eta_{i0}$, the condition (c) fails. Given the reference
state $\eta_{i0}$ alone, $Y_{it}(0)$ and the other post-treatment
outcomes remain correlated through the intervening states $(\eta_{i1},\ldots,\eta_{i,T_{\text{post}}})$.
\item With $U_{i}=(\eta_{i0},\eta_{it})$, the condition (c) is restored,
but completeness fails. By the Markov property, $\boldsymbol{Y}^{\text{pre}}_{i}\perp\eta_{it}\,|\,\eta_{i0}$,
so the proxy $\boldsymbol{Y}^{\text{pre}}_{i}$ carries no information
about $\eta_{it}$ and does not satisfy a completeness condition for
$(\eta_{i0},\eta_{it})$. The post-treatment outcomes after period
$t$ likewise carry no information about $\eta_{i0}$ given $\eta_{it}$.
\end{itemize}
Because no choice of $U_{i}$ satisfies the independence and the completeness
requirements at the same time, this mapping cannot identify the period-$t$
effect once the unobserved heterogeneity is time-varying.

This contrast isolates the contribution of the reference period outcome.
By treating $\boldsymbol{Y}^{\text{post}}_{i}(0)$ as a single measurement
rather than splitting it, and using the reference period outcome $Y_{i0}(0)$
as a third proxy that anchors identification on $\eta_{i0}$, this
paper identifies treatment effects at all post-treatment horizons
even when the latent state evolves.
\end{document}